\def\msol{{\rm\thinspace M${}_\odot$}}
\def\fig{{\rm\thinspace Figure}}

\def\kms{{km s${}^{-1}$}}
\def\etal{{\it et al.\thinspace}}
\def\eg{{\it e.g.\ }}

\def\ie{{\it i.e.\ }}

\def\msolyr{{\rm\thinspace M${}_\odot\; {\rm yr}^{-1}$}}
\def\gcm3{{g cm${}^{-3}$}}

\def\ga{{\rm\thinspace gauss}}

\def\msol{\hbox{$\rm\thinspace M_{\odot}$}}
\def\msolyr{\hbox{$\rm\thinspace M_{\odot} \; {\rm yr}^{-1}$}}

\def\h50{\hbox{$\rm\thinspace h_{50}$}}
\def\h50m1{\hbox{$\rm\thinspace h_{50}^{-1}$}}

\def\msol{\hbox{$\rm\thinspace M_{\odot}$}}
\def\etal{{\it et al.\thinspace}}
\def\eg{{\it e.g.\ }}

\def\ie{{\it i.e.\ }}

\def\fig{figure}
\def\Fig{Figure}

\def\p3m{P${}^3$M}
\def\ap3m{AP${}^3$M}
\def\-{{\em{---}}}

\documentclass{article}
\usepackage{emulateapj}

\begin{document}
\title{Implementing Feedback in Simulations of Galaxy Formation: \\
A Survey of Methods}

\author{R. J. Thacker}
\authoremail{thacker@astro.berkeley.edu}

\affil{Theoretical Physics Institute, Department of Physics, \\
University of Alberta, Edmonton, Alberta, T6G 2J1, Canada \\
and \\
Department of Physics and Astronomy, \\
University of Western Ontario, London, Ontario, N6A 3K7, Canada.\\
Current address\\
Department of Astronomy, \\
University of California at Berkeley,
Berkeley, CA, 94720.}
\and

\author{H. M. P. Couchman}
\authoremail{couchman@coho.physics.mcmaster.ca}
\affil{Department of Physics and Astronomy, \\
University of Western Ontario, London, Ontario, N6A 3K7, Canada.\\
Current address\\
Department of Physics and Astronomy, \\
McMaster University,
1280 Main St. West, Hamilton, Ontario, L8S 4M1, Canada.}

\begin{abstract}
We present a detailed investigation of a
number of different approaches to modelling feedback in simulations of
galaxy formation.  Gas-dynamic forces are evaluated using Smoothed
Particle Hydrodynamics (SPH). Star formation and supernova feedback are
included using a three parameter model which determines the star formation
rate (SFR)  normalization, feedback energy and lifetime of feedback
regions.  The star formation rate is calculated for all gas particles
which fall within prescribed temperature, density and convergent flow
criteria, and for cosmological simulations we also include a self-gravity
criterion for gas particles to prevent star formation at high redshifts. 
A Lagrangian Schmidt law is used to calculate the star formation rate from
the SPH density. Conversion of gas to stars is performed when the star
mass for a gas particle exceeds a certain limit, typically half that of
the gas particle. Feedback is incorporated by returning a precalculated
amount of energy to the ISM as thermal heating. We compare the effects of
distributing this energy over the smoothing scale or depositing it on a
single particle. Radiative losses are prevented from heated particles by
adjusting the density used in radiative cooling so that the decay of
energy occurs over a set half-life, or by turning off cooling completely
and allowing feedback regions a brief period of adiabatic expansion.  We
test the models on the formation of galaxies from cosmological initial
conditions and also on isolated disk galaxies. For isolated prototypes of
the Milky Way and the dwarf galaxy NGC 6503 we find feedback has a
significant effect, with some algorithms being capable of unbinding gas
from the dark matter halo (`blow-away'). As expected feedback has a
stronger effect on the dwarf galaxy, producing significant disk
evaporation and also larger feedback `bubbles' for the same parameters. In
the critical-density CDM cosmological simulations, evolved to a redshift
$z=1$, we find the reverse to be true. Further, feedback only manages to
produce a disk with a specific angular momentum value approximately twice
that of the run with no feedback, the disk thus has an specific angular
momentum value that is characteristic of observed elliptical galaxies.  We
argue that this is a result of the extreme central concentration of the
dark halos in the standard CDM model and the pervasiveness of the
core-halo
angular momentum transport mechanism (even in light of feedback). A
simulation with extremely violent feedback, relative to our fiducial
models, leads to a disk that resembles the other simulations at $z=1$ and
has a specific angular momentum value that is more typical of observed
disk galaxies.  At later times, $z=0.5$, a large amount of halo gas which
does not suffer an angular momentum deficit is present, however the
cooling time is too long to accrete on to the disk.  We further point out
that the disks formed in hierarchical simulations are partially a
numerical artifact produced by the minimum mass scale of the simulation
acting as a highly efficient `support' mechanism. Disk formation is
strongly affected by the treatment of dense regions in the SPH method. The
problems inherent in the treatment of high density regions in SPH, in
concert with the difficulty of representing the hierarchicial formation
process, means that realistic simulations of galaxy formation require far
higher particle resolution than currently used.

\end{abstract}
\section{Introduction}
A detailed understanding of galaxy formation in hierarchically clustering
universes remains one of the primary goals of modern cosmology. Whereas on
large scales the clustering of matter is determined almost solely by
gravitational forces, a large number of
physical processes contribute to galaxy formation. Further complication is
evident in that a significant number of these processes cannot be modelled
from first principles in a simulation of galaxy formation, the main
example of this being star formation. 

Analytic and semi-analytic theories of galaxy formation are well-developed
(see White 1994, for an overview). The theoretical framework for studying
the condensation of baryons in dark matter halos was laid out by White \&
Rees (1978) following the foundational work on hierarchical clustering by
Peebles (1980, and references therein). White and Rees illuminated the
fact that for baryons condensing in dark matter halos of sub-galactic
($\ga 10^6$ \msol) size the cooling time of the gas is always shorter than
the free-fall collapse time. In a related paper, Fall \& Efstathiou (1980) 
demonstrated that disk galaxies could be formed in a hierarchical
clustering model, provided that the gas maintains its angular momentum. 
Later work by White \& Frenk (1991)  further developed this hierarchical
clustering model and identified the {\em cooling catastrophe} for CDM
cosmologies.  The cooling catastrophe is caused by the large mass fraction
in small high-density halos at early times. The gas resident in these
halos cools on the time-scale of a Myr, thus precipitating massive star
formation very early on in the development of the CDM cosmology (at odds
with the observations of our Universe).  To circumvent this problem White
\& Frenk introduced star formation and the associated feedback from
supernovae and showed that, given plausible assumptions, the cooling
catastrophe can be avoided. The main deficiency in the semi-analytic
programmes is that they cannot describe the geometry of mergers which is
exceptionally important in the assembly of galaxies. 
 
Smoothed Particle Hydrodynamic simulations of galaxy formation
(\cite{nk,NW94,E94}, for example)  detail the hierarchical merging
history, but have been limited in terms of resolution.  Achieving high
resolution is particularly difficult. Since galaxy formation is affected
by long range tidal forces, the simulation must be large enough to include
these, which in turn enforces a low mass resolution. Two solutions to this
problem exist; the first samples the long range fields using lower
particle resolution than the main simulation region (the multiple mass
technique, Porter 1985) while the second includes the long
range fields as a pre-calculated low resolution external field (\eg
\cite{SL98}).  Simulations performed in this manner represent lengths
scales from 50 Mpc to 1 kpc, a dynamic range of $5\times10^4$.

It is well known that in SPH simulations it is comparatively easy to form
flattened disk structures resembling disk galaxies. However, the resulting
gaseous structures are deficient in angular momentum (see Navarro, Frenk
\& White 1995, for a comparison of a number of simulations to observed
galaxies). The loss of angular momentum occurs during the merger process
as dense gas cores lose angular momentum to the dark matter halos (see
Barnes 1992, for an explanation of the mechanism). This is a very
significant problem since a fundamental requirement of the disk formation
model, presented in Fall \& Efstathiou (1980), is that the gas must
maintain a similar specific angular momentum to the dark matter to form a
disk.  The solution to this problem is widely believed to be the inclusion
of star formation and feedback from supernovae and stellar winds.  By
including these effects the gas should be kept in a more diffuse state
which consequently does not suffer from the core-halo angular momentum
transport problem.  Notably, a recent letter by Dominguez-Tenreiro,
Tissera \& Saiz (1998), based upon analytic work by Christodoulou,
Shlosman \& Tohline (1995) and van den Bosch (1998), has shown that the
inclusion of star formation may go some way in helping to resolve the
angular momentum problem. They suggest that a second mechanism, bar
formation, also contributes to the loss of angular momentum within the
disk. Including a photoionizing background (Efstathiou 1992)  appears to
make the problem slightly less severe (Sommer-Larsen, Gelato \& Vedel
1998) but this is the subject of debate; for a conflicting viewpoint see
Navarro \& Steinmetz (1997). 

Individual supernovae cannot be modelled in galaxy formation simulations.
Consequently, there is no {\em a priori} theory for how feedback energy
should be distributed in the simulation. A decision must be made whether
feedback energy should be passed to the thermal or kinetic sector of the
simulation.  As a direct result of this `freedom', a number of different
algorithms for the distribution of feedback energy have been proposed
(\cite{nk,NW93,MH94} for example). Since the interstellar medium is
multi-phase (\cite{OM77}), the feedback process should ideally be
represented as an evaporation of molecular clouds.  Note that in a
single-phase model, the analogue of the cloud evaporation process is
thermal heating. A seminal attempt at representing this process has been
made by Yepes \etal (1997), and recently Hultman and Pharasyn (1999)  have
adapted the Yepes \etal model to SPH. At the moment it is unclear how well
motivated these models are since (1) they must make a number of
assumptions about the physics of the ISM and (2) they are not truly
multiphase, since the dynamics are still treated using a single phase. 
Consequently, given the uncertainties inherent in multiphase modeling, the
investigation presented here explores a single phase model.

As a first approximation, the star formation algorithms can be divided
into three groups. The first group contains algorithms which rely upon
experimental laws to derive the star formation rate (\cite{MH94}, for
example). The second group contains those which predict the SFR from
physical criteria (\cite{nk}, for example). The third group is comprised
of algorithms which do not attempt to predict the SFR, but instead set a
density criterion for the gas so that when this limit is reached the gas
is converted into stars (\cite{GI97}, for example). In this work the
first approach is taken.

Given that feedback occurs on sub-resolution scales, it is difficult to
decide upon the scale over which energy should be returned. However, SPH
incorporates a minimum scale automatically\-the smoothing scale. Katz
(1992) was the first to show that simply returning a specified amount of
energy to the ISM is ineffective: the characteristic time-scale of
radiative cooling at high density is far shorter than the simulation
dynamical time-scale. This is, at least partially, another drawback of
trying to model sub-resolution physics at the edge of the resolution
scale. Further, the strong disparity between time-scales means that not
only are length scales sub-resolution, {\em so is the characteristic
temporal evolution}. This point has been made several times in relation to
simulations with cooling (\eg Katz 1992) but it seems even more important
in the context of models with feedback. Thus, in an attempt to circumvent
this problem, a new thermal feedback model is presented in which the
radiative losses are reduced by changing the density used in the radiative
cooling equation. The density used is predicted from the ideal gas
equation of state and then integrated forward, decaying back to the SPH
density (in a prescribed period). The effect of preventing radiative
losses using a brief adiabatic period of evolution for the feedback region
(\cite{ger})  is also examined. An alternative method of returning thermal
energy is to heat an individual SPH particle (\cite{ger}). This method has
been shown to be effective in simulations of isolated dwarf galaxies. The
final mechanism considered is one that attempts to increase the energy
input from SN to account for the high radiative losses. Mechanical
feedback boosts are not considered for the following reasons; (1)
parameters that are physically motivated (\eg feedback efficiency of 10\%)
produce too much velocity dispersion, and (2)  the method makes no account
for force softening.

The response of the (simulated) ISM to a single feedback event is examined
to gain an understanding of the qualitative and quantitative performance
of each feedback algorithm.  To determine whether the parameters of the
star formation are more important than the dynamics of the simulation, an
exploration of the parameter space of one of the algorithms is undertaken.
Since feedback is expected to have a more significant effect on dwarf
systems (due to the lower escape velocity) the effect of the feedback
algorithms on a Milky Way prototype is contrasted against a model of the
dwarf galaxy NGC 6503. Following this investigation, a series of
cosmological simulations is conducted to examine whether the conclusions
from the isolated simulations hold in a cosmological environment.
Particular attention is paid to the rotation curves and angular momentum
transport between the galaxy and halo.

The structure of the paper is as follows:  In section~\ref{nummeth},
important features of the numerical technique are reviewed.
 In sections~\ref{sfrpresc}-\ref{res}, the star formation
prescription is presented and a detailed analysis of its
performance on isolated test objects presented conducted. Results
from simulations are reviewed in section \ref{res}, and a summary of the
findings is given in section~\ref{conc2}. 

\section{Numerical Method}\label{nummeth}
An explicit account of the numerical method, including the equation of
motion and artificial viscosity used, is presented in Thacker \etal
(1998). The code is a significant development of the publically available
HYDRA algorithm (Couchman, Thomas \& Pearce 1995). The main features of
the method are summarized for clarity. Gravitational forces are evaluated
using the adaptive Particle-Particle, Particle-Mesh algorithm (AP${}^3$M,
\cite{HC91}).  Hydrodynamic evolution is calculated using SPH. Notable
features of algorithm relevant to the hierarchical simulations follow,

\begin{itemize}
\item The neighbour smoothing is set to attempt to smooth
over 52 neighbour particles, usually leading to a particle having between
30 and 80 neighbours. This number of neighbors assures a stable
integration and reduces concern over the relative fluctuation
in neighbor counts.

\item The minimum hydrodynamic resolution
scale is set by $h_{min}=\epsilon/2$ where $\epsilon$ is the
gravitational softening. If the minimum gravitational resolution is
$2\epsilon$ and the minimum hydrodynamic resolution $4h_{min}$ then the
two of them match with this definition of $h_{min}$.

\item Once the smoothing length of a particle reaches $h_{min}$, {\em all}
particles within
$2h_{min}$ are smoothed over. Thus at this scale the code changes from an
adaptive to nonadaptive scheme. This avoids mismatched gravitational and
hydrodynamic forces at scales close to that of the resolution
(\cite{BB97}).

\end{itemize}
Radiative cooling is implemented using an assumed 5\% $Z_\odot$
metallicity. The precise cooling table is interpolated from Sutherland and
Dopita (1993). Radiative cooling is calculated in the same fashion as
discussed in Thomas \& Couchman (1992). 

\section{Star Formation Prescription}\label{sfrpresc}
\subsection{Implementation details}
The star formation algorithm is based on that presented in Mihos and
Hernquist (1994). Kennicutt (1998) has presented a
strong argument that the Schmidt Law, with star formation index
$\alpha=1.4\pm 0.15$, is an excellent model. It characterizes star
formation over {\em five decades of gas density}, although it does exhibit
significant scatter.

For computational efficiency, a Lagrangian version of the
Schmidt Law is used, that corresponds to a star formation index
$\alpha=1.5$,
\begin{equation}
\label{sfr} {d M_* \over d t} = C_{\it sfr}\; \rho_{g}^{1/2}
\; M_{g}, 
\end{equation} 
where $C_{\it sfr}$ is the star formation rate normalization, the $g$
subscripts denotes gas and the $*$ subscript stars.  Assuming
approximately constant volume over a time-step, both sides may divided
by the volume and the standard Schmidt Law with index
$\alpha=1.5$ is recovered. A value of $\alpha=1.5$ is preferred since it
leads to the
square root of the gas density, which is numerically efficient
and the value is within the error bounds. As written the constant
$C_{\it sfr}$ has units of $\rho^{-1/2}t$, but can be made
dimensionless by multiplying by $(4\pi G)^{1/2}$. Hence, equation
\ref{sfr}
may be written with dimensionless constants, leading to the following
form,
\begin{equation} 
{d M_* \over d t} = \sqrt{4\pi G} \, c^* \; 
\rho_{g}^{1/2} \;  M_{g} = \frac{c^*M_g}{t_{ff}}, 
\end{equation} 
where $c^*$ is the dimensionless star formation rate (\cite{nk}). 
The range for this parameter is reviewed in section \ref{tests}. Limiting
the star formation rate by the local cooling time-scale was not
considered.
The reason for this is that typically the cold dense cores which form
stars are at the effective
temperature minimum of the simulation for non-void regions, namely $10^4$
K, which corresponds to the end of the radiative cooling curve.

Star formation is allowed to proceed in regions that satisfy the following
criteria,
\begin{enumerate}
\item the gas exceeds the density limit of $2\times10^{-26}
\rm{g\;cm}^{-3}$
\item the flow is convergent, ($\nabla.{\bf v}<0$)
\item the gas temperature is less than $3\times10^4$ K
\item the gas is partially
self-gravitating,
$\rho_{gas}>0.4\rho_{dm}$
\end{enumerate}

The first criterion associates star formation with dense regions
(regardless of the underlying dark matter structure). The
second criterion is included to link star formation with regions that are
collapsing. The third prevents star formation from occurring in regions
where the average gas temperature is too high for star formation. The
final criterion is particularly relevant to cosmological simulations since
it limits star formation to regions where the dynamics are at least
partially determined by the baryon density. Since the mass scales probed
by the simulation are considerably larger than the mass scales of
self-gravitating cores in which star formation would occur, we use a
pre-factor of 0.4 to help compensate for this disparity.  

Representing the growth of the stellar component of the simulation
requires compromises. It is clearly impossible to add particles with
masses $dM_*$ at each time-step since this would lead to many millions of
small particles being added to the simulation. Alternatively, a gas
particle may be viewed as having a fractional stellar mass component, \ie
the particle is a gas-star hybrid (in the terminology of \cite{MH94}).
Thus, as gas is turned into stars, the stellar mass increases while the
gas mass decreases. Mihos and Hernquist take this idea to its limit by
only calculating gas forces using the gas mass of a particle
(gravitational forces are unaffected because of mass conservation). A
drawback of this method is that it still enforces a collisional trajectory
on the collisionless stellar content of a particle. While this is a good
description of real physics, since stars take a number of galactic
rotations to depart from the gas cloud in which they are born, it is in
strong disagreement with what would happen to collisional and
collisionless particles in a simulation.  To decide when to form stars,
Katz (1992) and Steinmetz \& Muller (1994, 1995) use a star formation
efficiency.  Once the star mass of the gas particle reaches a set
(efficiency)  percentage of the gas particle mass then a star particle is
created with that mass value. This also leads to the potential spawning of
very many gas particles.  As a compromise, the scheme used in this work is
as follows:  Once $\dot{M}_*$ has been evaluated, the associated mass
increase over the time-step is added to the `star mass' of the hybrid
gas-star particle. Note, in this model the `star mass' is a `ghost'
variable and does not affect dynamics in any way, \ie until the star
particle is created the particle is treated as entirely gaseous when
calculating hydrodynamic quantities. Two star particles (of equal mass)
can be created for every gas particle. This ensures that feedback events
occur frequently since we do not have to wait for the entire gas content
of a region to be consumed before feedback occurs, and at the same time
prevents the spawning of too many star particles. The creation of a star
particle occurs when the star mass of a gas-star particle reaches one half
that of the mass of the {\em{initial}} gas particles. The gas-star
particle mass is then decremented accordingly. The second star particle is
created when the star mass reaches 80\% of half the initial gas mass. SPH
forces are calculated using the total mass of the hybrid particles.  Note,
in all the following sections, ``gas particle'' should be interpreted
meaning ``gas-star hybrid particle''. These assumptions yield a star
formation algorithm where each gas particle has a star formation history. 
This assumption is motivated because cloud complexes in any galaxy have an
associated star formation history.

There are two notable drawbacks to this algorithm. Firstly, since star
particles are only formed when the star mass exceeds a certain threshold
there is a delay in forming in stars. As a consequence, prior to the star
particle being formed, the SPH density used in the calculation of the SFR
will be overestimated. By selecting the star mass to be one half that of
the gas mass this problem is reduced, but it is not removed. Secondly,
until the first star particle is spawned, the trajectory of the stellar
component is entirely determined by the gas dynamics.

Once a star particle is created the associated feedback must be evaluated. 
A simple prescription is utilized, namely that for every $100 M_\odot$ of
stars formed there is one supernovae which contributes $10^{51}$ erg to
the ISM. This value is used in Sommer-Larsen \etal
(1998), and feeds back $5\times 10^{15}$ erg
g${}^{-1}$ of star particle to the surrounding ISM. Since this
value is a specific energy, a temperature can be associated with feedback
regions (see section~\ref{rfe}).  Navarro \& White (1993), using a
Salpeter IMF, with power law slope 1.5 and mass cut-offs at 0.1 and 40
$M_\odot$, derive that $2\times10^{15}$ erg g${}^{-1}$ of star particle
created is fed back to the surrounding gas.  The actual value is subject
to the IMF, but scaling between values can be achieved using a single
parameter which we label $e^*$. For $e^*=1$ the energy return is $5\times
10^{15}$ erg
g${}^{-1}$, while $e^*=0.4$ gives the Navarro \& White value. Only brief
attention is paid to changing this parameter since
it is more constrained than the others in the model.  Variations in
metallicity caused by the feedback process (\cite{MS98}) are not
considered, as this involves a complicated feedback loop involving the gas
density and cooling rate. 

\subsection{Energy feedback}\label{rfe}
The following sections describe each of the feedback algorithms
considered.

\subsubsection{Energy smoothing (ES)}
The first of the methods is comparatively standard:  the total feedback
energy is returned to the local gas particles. For simplicity in the
following argument, it is assumed a tophat kernel is used to feedback the
energy over the nearest neighbour particles. The number of neighbours is
constrained to be between 30 to 80 which may divorce the feedback
scale from the minimum SPH smoothing scale set by $h_{min}$. This occurs
in the densest regions where the average interparticle spacing becomes
significantly less than $h_{min}$. This actually renders feedback more
effective in these regions than it would be if all the energy were
returned to the particles with in $2h_{min}$ region, but the effect is not
significant.  Given these
assumptions for star formation, and working in units of
internal energy,
it
is possible to evaluate the temperature increase, $\Delta T$ of a feedback
region as follows,
\begin{equation}
\Delta T={2 \over 3}{\mu m_p \over  k} {E_{SN} M_* \over N_s M_g},
\end{equation} 
which given $N_s=52$, $E_{SN}=5\times10^{15} \; \rm{erg} \; \rm{g}^{-1}$, 
and $M_*/M_g=1/2$ yields,
\begin{equation}
\Delta T \simeq 2.4 \times 10^7 {M_*\over N_s M_g}\simeq 2.3\times 
10^5
\;
\rm{K}.
\end{equation}
Clearly this boost may be increased or decreased by altering any one of
the variables $E_{SN}$, $N_s$ and the ratio $M_*/M_g$. Keeping $E_{SN}$
constant, but reducing $N_s$ to 32 and increasing $M_*/M_g$ to 1 would
yield $\Delta T \simeq 7\times 10^5 \; \rm{K}$, demonstrating the
sensitivity of feedback to the SPH smoothing scale. 

\subsubsection{Single particle feedback (SP)}
Alternatively, all of the energy may be returned to a single SPH particle.
Gerritsen (1997) has shown that this is an effective prescription in
simulations of evolved galaxies, yielding accurate morphology and physical
parameters. In this case the temperature boost is trivially seen to be
\begin{equation}
\Delta T \simeq 2.4 \times 10^7 \rm{K}.
\end{equation}
as $N_s=1$ and $M_*/M_g=1$. There is one minor problem in
that when the gas supply is exhausted, the mechanism has no way of
returning the energy (unless one continues to make star particles of
smaller and smaller masses). As a compromise the nearest SPH
particle is found and the energy is given to this particle.

\subsubsection{Temperature smoothing (TS)}
The final feedback mechanism considered is one that accounts for the fast
radiative losses by increasing the energy input. The first step is to
calculate the temperature a single particle would have if all the energy
were returned to it (as in the SP model).  Then this value is smoothed
over the local particles using the SPH kernel. This method leads to a
vastly higher energy input than the others and represents a case of
extreme feedback (essentially 52 times the ES feedback). 
For the isolated simulations, the cooling
mechanism (see below) was not adjusted in this model since the feedback
regions in the disk have cooling time of several time-steps
(at least 10). In the cosmological simulations, the
alternative cooling mechanisms were considered since the cooling time is
only a few time-steps (\cite{nk,SL98}).

\subsubsection{Preventing immediate radiative energy losses}
As was previously discussed, Katz (1992) was the first to show that
feedback energy returned to the ISM is radiated away extremely quickly in
high density regions. This is a result of the characterisitic
dynamical time of the
simulation being far longer than that of cooling.

The first method for preventing the immediate radiative loss of the
feedback energy is to alter the density value used in the radiative
cooling mechanism. This change is motivated by Gerritsen's (1997) tests on
turning off radiative cooling in regions undergoing feedback. Assuming
pressure equilibrium between the ISM phases (which is not true after a SN
shell explodes but is a good starting point) one may derive the estimated
density that the region would have after the SN shell has exploded. If the
local gas energy is increased by $E_{SN}$, then the perfect gas equation
of state yields
\begin{equation}
\rho_{est}={  E_{i} \rho_{i} \over E_{i}+E_{SN}}. 
\end{equation}
Following a feedback event the estimated density is allowed to
decay back to it's local SPH value with a half-life $t_{1/2}$. 
{\epsscale{0.95} \plotone{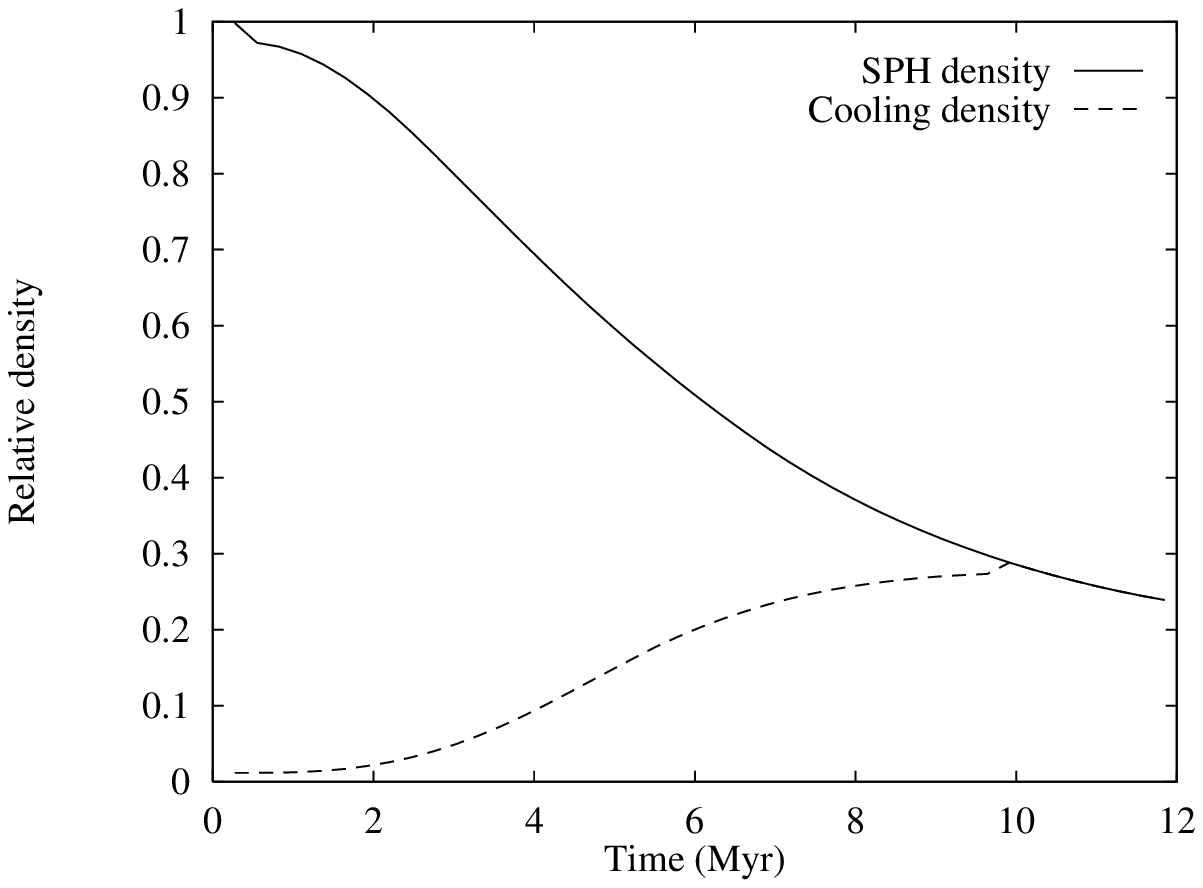}}

{\small {\footnotesize {\sc Fig.}~1.}---Evolution of the cooling density
and
the
SPH
density following a single
feedback event in the ESna scheme. $t_{1/2}=5$ Myr, and clearly the values
converge within $2t_{1/2}$. Note the small initial drop in the SPH density
in response to the feedback energy, followed by a slower expansion. The   
initial cooling density is approximately 5\% of the SPH density,
consistent with the feedback temperature of $2\times10^5$ K in an ambient
10,000 K region.}
\bigskip

\noindent To
calculate the decay rate (\ie to predict the cooling density at
the time-step $n+1$ from that at time-step $n$) the following
function is used
\begin{equation}
\footnotesize
\rho_{cool}^{n+1}=\cases{
\rho_{cool}^n + dt\times\Delta\rho^n (t_f^2/ t_{1/2}^3) e^{-0.33
(t_f/t_{1/2})^3}, & $t_f \leq t_{1/2}$; \cr
\noalign{\vspace{5pt}}
\rho_{SPH}^n - \Delta\rho^ne^{-0.693dt/t_{1/2}},  &
$t_{1/2}\leq t_f<3t_{1/2}$; \cr
\noalign{\vspace{5pt}}
\rho_{SPH}^{n+1}, & $ 3t_{1/2} \leq t_f$ ,
}
\end{equation}
where $t_f$ is time since the feedback event occurred,
$\Delta\rho^n=\rho_{SPH}^n-\rho_{cool}^n$, and $dt$ is the time-step
increment. Once a region passes beyond $3t_{1/2}$, the
cooling density is forced to be equivalent to the SPH density, although
usually the
values converge within $2t_{1/2}$. This comparatively complex
function was chosen because in a simple exponential decay model, the
cooling
density 
increases by the largest amount immediately following the feedback
event. To have any effect the cooling density must be allowed
to persist at its low value for a reasonable period of time. In
\fig~1
the two densities calculated after a single feedback event are compared.
This cooling mechanism is denoted by a suffix {\em na} (for
non-adiabatic) on the energy input acronym. Springel and White (in prep)
have considered a similar model (Pearce, private communication).

\setcounter{figure}{1}
\begin{figure*}[t]
\epsfxsize=15cm
\centerline{\epsffile{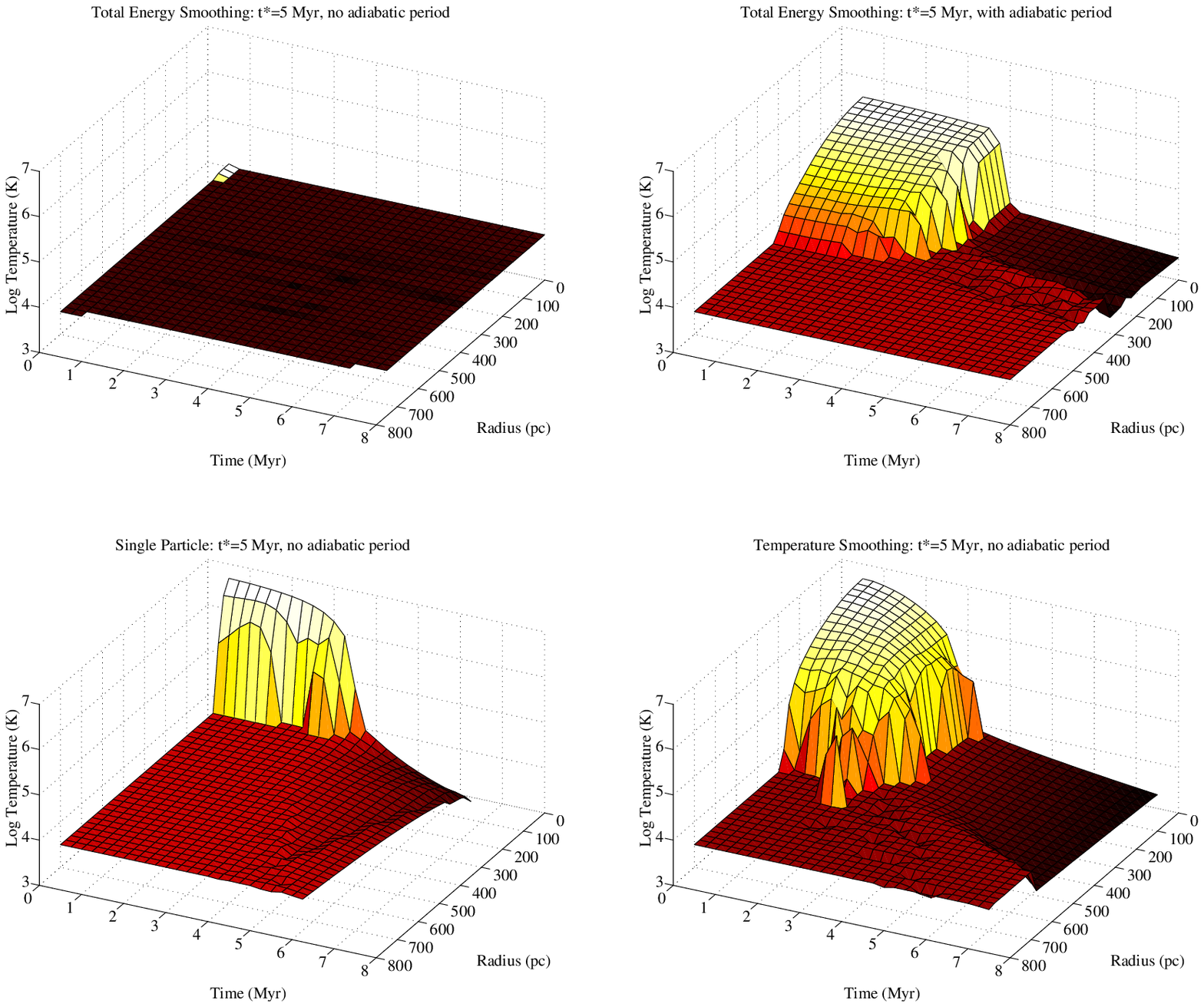}}
\caption{Evolution of the cooling density and the SPH
density following a single feedback event in the ESna scheme. $t_{1/2}=5$
Myr, and clearly the values converge within $2t_{1/2}$. Note the small   
initial drop in the SPH density in response to the feedback energy,      
followed by a slower expansion. The initial cooling density is
approximately 5\% of the SPH density, consistent with the feedback
temperature of $2\times10^5$ K in an ambient 10,000 K region.}
\label{den.comp}
\end{figure*}   

Gerritsen allowed his SPH particles to remain adiabatic for $3\times10^7$
years, approximately the lifetime of a $8 M_{\odot}$ star. It is difficult
to argue what this value should be and hence the $t_{1/2}$ parameter space
was explored. Note that if during the decay period another feedback event
occurs in the local region, the density value used is the minimum of the
current decaying one and the new calculated value.

As a second method for preventing radiative losses Gerritsen's idea was
utilized: make the feedback region adiabatic. This is easily achieved in
our code by using the same mechanism that calculates the estimated density
value. Provided the estimated density value is less than half that of the
local SPH value then the particle is treated as adiabatic. Above this
value the estimated density is used in the radiative cooling equation. 
This cooling mechanism is denoted with a suffix {\em a}. 

\subsection{Comparison of methods} 
To test each of the feedback methods
and gain insight into their effect on the local ISM, a single feedback
event was set up within a prototype isolated Milky Way galaxy. The
evolution of the particles within $3h$ of the feedback event were
followed. The time evolution of the temperature versus radius
for each scheme is shown in \fig~2. 

\subsubsection{Qualitative discussion}\label{profiles}

The adjusted cooling mechanism has little effect on the ES run because the 
estimated density is not low enough to increase the cooling time
significantly beyond the length of the time-step. Including a prefactor of
0.1 in the equation, so that the estimated density is significantly lower,
does increase the cooling time sufficiently. However, since introducing
the adiabatic phase allows the feedback energy to induce expansion, we
prefer this method, rather than trying to adjust the estimated density
method. For the SP run the density reduction is much higher
(since the total energy, $E_{SN}$, is applied to the single particle) and 
hence the mechanism does allow the particle to remain hot. There is 
little perceptible difference between the SPa and SPna profiles.

Both SP feedback, TS, and ESa induce noticeable expansion
of the feedback region. Thus after the heat input has been radiated away,
the continued expansion introduces adiabatic cooling (since the temperature
of the region falls below the $10,000$ K cut off of the cooling curve). It
is particular noticeable in the TS plot where the low temperature plateau
continues to widen quite drastically and this is manifest in the
simulation
with the appearance of a large bubble in the disk. Caution should
be exercised in
interpreting these bubbles in any physical manner, since their size
is set solely by the resolution scale of the SPH. ESa also produces
bubbles, but due to the lower temperature there is less expansion. Single
particle feedback produces the smallest bubbles and often ejects the hot
particle vertically from the disk. This occurs with regularity since the
smoothing scale is typically larger than the disk thickness: if the hot
particle resides close to the edge of the disk the pressure forces from
the surrounding particles will be asymmetric resulting in a strong `push'
out of the disk.  Note that this is phenomenologically similar to the
mechanism by which SN gas is ejected from disks (\cite{OM77}), although
this
should not be over-interpreted. 

The only scheme which stands out in this investigation is ESna: the
cooling mechanism fails to prevent drastic radiative losses. The remainder
of the algorithms produce an effect on both the thermal
properties of the ISM and its physical distribution.

\subsubsection{Effect of methods on time-step criterion}
Since our code does not have multiple time-steps, it is important to
discern whether one method allows longer time-steps than another. This is
a desirable feature since it reduces the wall-clock time for simulations.
Of course an algorithm which has a fast wall-clock time but produces poor
results would never be chosen, however for two algorithms with similar
results this criterion provides a useful parameter to choose one over the
other. A comparison of SPna, TS, ESna and the no
feedback (NF) run is displayed in \fig~3.

The simulation time versus the number of time-steps was compared for data
from the Milky Way prototype runs (see section~\ref{mw}). The SP methods
produce the shortest time-step, requiring almost double the number of
time-steps than the NF model.  The ES variants require only 10\% more
time-steps than the run without feedback. In SP feedback the acceleration
felt by the hot particle limits the time-step significantly.  TS also
requires more time-steps than ES due to the rapid expansion of feedback
regions. Typically though, the number of time-steps required is somewhat
less (20\%) than that for single particle feedback. 

\subsection{Explored parameter space of the algorithm}\label{tests}
All models exhibit dependencies on the free parameters $c^*$ and $e^*$,
corresponding to the SFR normalization and efficiency of the feedback
energy return. The models which use a modified cooling formalism
also exhibit a dependence upon $t_{1/2}$, the approximate half-life of
feedback regions. To simplify matters, an ensemble
with $e^*=0.4$ (the value used in Navarro \& White 1993) was run, allowing
us to concentrate on the effect of the $c^*$ and $t_{1/2}$.
To determine the effect of varying $e^*$, two more
simulations with $e^*=1$ were run.

\begin{center}
\begin{tabular}{clccc}
\hline\hline
Run & $c^*$ & $t_{1/2}^{\rm a}$ & $e^*$ &
$N_{step}^{\rm b}$ \\
\hline
5001 & 0.001  & 0   & 0.4 & 1999  \\
5002 & 0.003  & 0   & 0.4 & 2001  \\
5003 & 0.01   & 0   & 0.4 & 1924  \\
5004 & 0.03   & 0   & 0.4 & 1977  \\   
5005 & 0.1    & 0   & 0.4 & 1989  \\
5006 & 0.3    & 0   & 0.4 & 2140  \\
5007 & 1.0    & 0   & 0.4 & 2087  \\
7001 & 0.033   & 1.  & 0.4 & 1989  \\
7002 & 0.033   & 5.  & 0.4 & 2000  \\
7003 & 0.033   & 10. & 1.0 & 1947  \\
7004 & 0.033   & 1.  & 1.0 & 1938  \\
7005 & 0.033   & 10. & 0.4 & 1940  \\
\hline
\end{tabular}
\end{center}
\medskip
\noindent{\small \bf Table 1:}
{\small Summary
of star formation parameter space simulations. The simulations
were of a rotating cloud collapse (see \cite{rob}). ${}^{\rm
a}$ $t_{1/2}=0$ denotes that feedback was removed from the
simulation. ${}^{\rm b}$ The number of steps are given to t=1.13, the
final point of the parameter space plot.} 
\bigskip

\subsubsection{Simple collapse test}
To gain an understanding of the algorithms in a simple collapse model
(that also may be run in a short wall-clock time) the rotating cloud
collapse model of Navarro and White (1993, also see \cite{rob})
was utilized. Such models
actually bear little resemblance to the hierarchical formation picture,
but they do allow a fast exploration of the parameter space. 

For this test the self-gravity requirement was removed. The reason for
this is that in cosmological simulations it is virtually guaranteed that
the gas in a compact disk will be self-gravitating. This is due to the low
number of dark matter particles in the core of the halo relative to the
number of gas particles.

The most important parameter in the star formation model is the $c^*$
parameter since it governs the SFR normalization. Therefore, an ensemble
of models was run with $c^*\in[0.001,1]$ (and $e^*=0.4$). The secondary
parameter in the model, $t_{1/2}$, is expected to have comparatively
little effect on the star formation rate (due to the low volume factor of
regions undergoing feedback). Hence only a range of plausible alternatives
were considered, namely $t_{1/2}=1,5,10$ Myr, in the ESa model. The
simulation parameters are detailed in table~1. 

\subsubsection{Results}
\Fig~4 displays a plot of the $c^*$ parameter space, showing SFR and $c^*$
versus time. Feedback was effectively turned off in this simulation by
reducing the energy return efficiency to $10^{-7}$. Although a severe
amount of smoothing has had to be applied (a running average over 40
time-steps, followed by linear interpolation on to the grid) there are a
number of interesting results. 

The time at which the peak SFR occurs is almost constant across all values
of $c^*$. This is an encouraging result since it indicates that the time
at which star formation peaks is dictated by dynamics and not by the
parameters of the model (at least without feedback). In fact the SFR peak
time corresponds to the time when the collapse model reaches its highest
density, following this moment a significant amount of relaxation occurs
and the gas has a \\

{\epsscale{0.95} \plotone{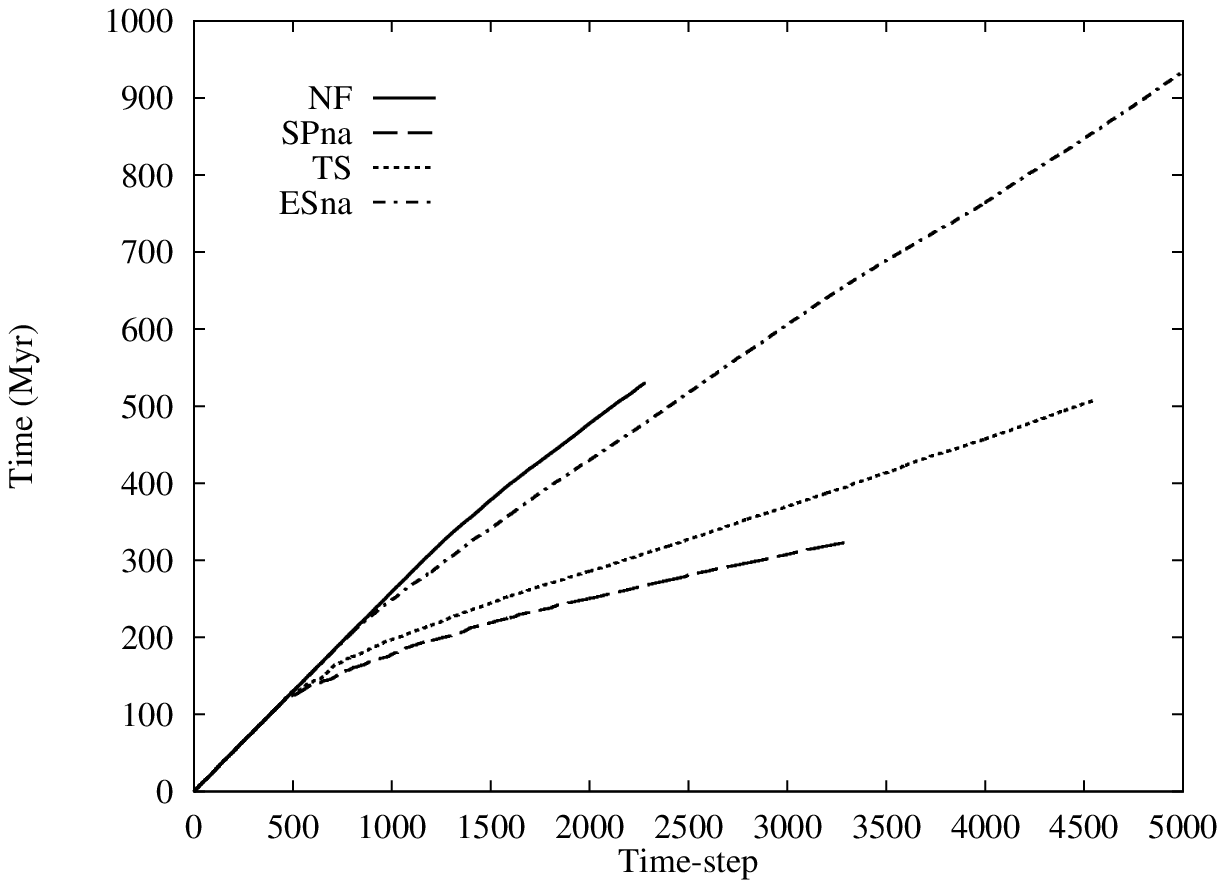}}

{\small {\footnotesize {\sc Fig.}~3.}---Effect of feedback scheme on the
time-step
selection in the
simulation. Data from the Milky Way prototype runs is shown. Twice
as many time-steps are required for SP feedback compared to runs  
without feedback. The ESa and SPa runs
are not shown, but are approximately 5\% lower than the
respective runs without the adiabatic period.}
\bigskip

{\epsscale{0.95} \plotone{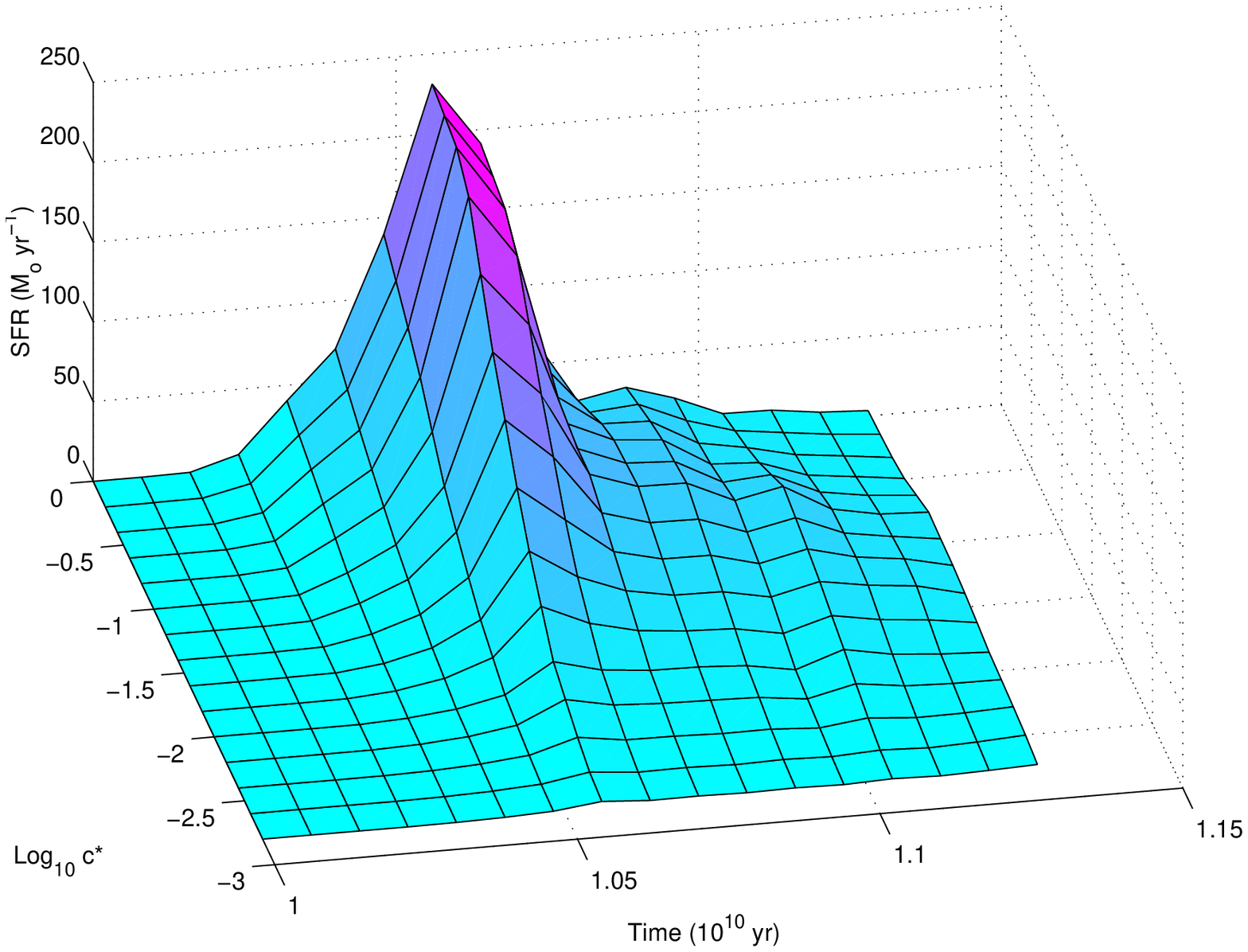}}

{\small {\footnotesize{\sc Fig.}~4.}---Dependence of
the SFR on the $c^*$ parameter in a
model with no feedback. The data for seven runs was linearly interpolated
to form the plotted surface. The time of the peak SFR moves very little
with changing $c^*$, and almost all the models can be fitted to
exponential
decay models following the peak SFR epoch.}
\bigskip

\noindent lower average density. Note, however, that this is an idealized
model with no feedback and a uniform collapse. 

\Fig~5 shows the dependence of the SFR on the $t_{1/2}$ and $e^*$
parameters. To detect trends in the SFR, a running average is shown,
calculated over 40 time-steps. Clearly, there is little distinction
between the runs with $t_{1/2}=1$ and 10. This can be attributed to the
low volume fraction of regions undergoing feedback. It is interesting to
note that the SFR is more sensitive to the amount of energy returned,
dictated by the $e^*$ parameter, than the lifetime for which this energy
is allowed to persist. The line corresponding to $e^*=1$ (the standard
energy return value of $5\times10^{15} \;{\rm erg}\;{\rm g}^{-1}$) does
not have the secondary and tertiary peaks in the SFR exhibited by the
$e^*=0.4$ runs.  This is probably attributable to the $\nabla.{\bf v}$
criterion: the $e^*=1$ \\
{\epsscale{0.95} \plotone{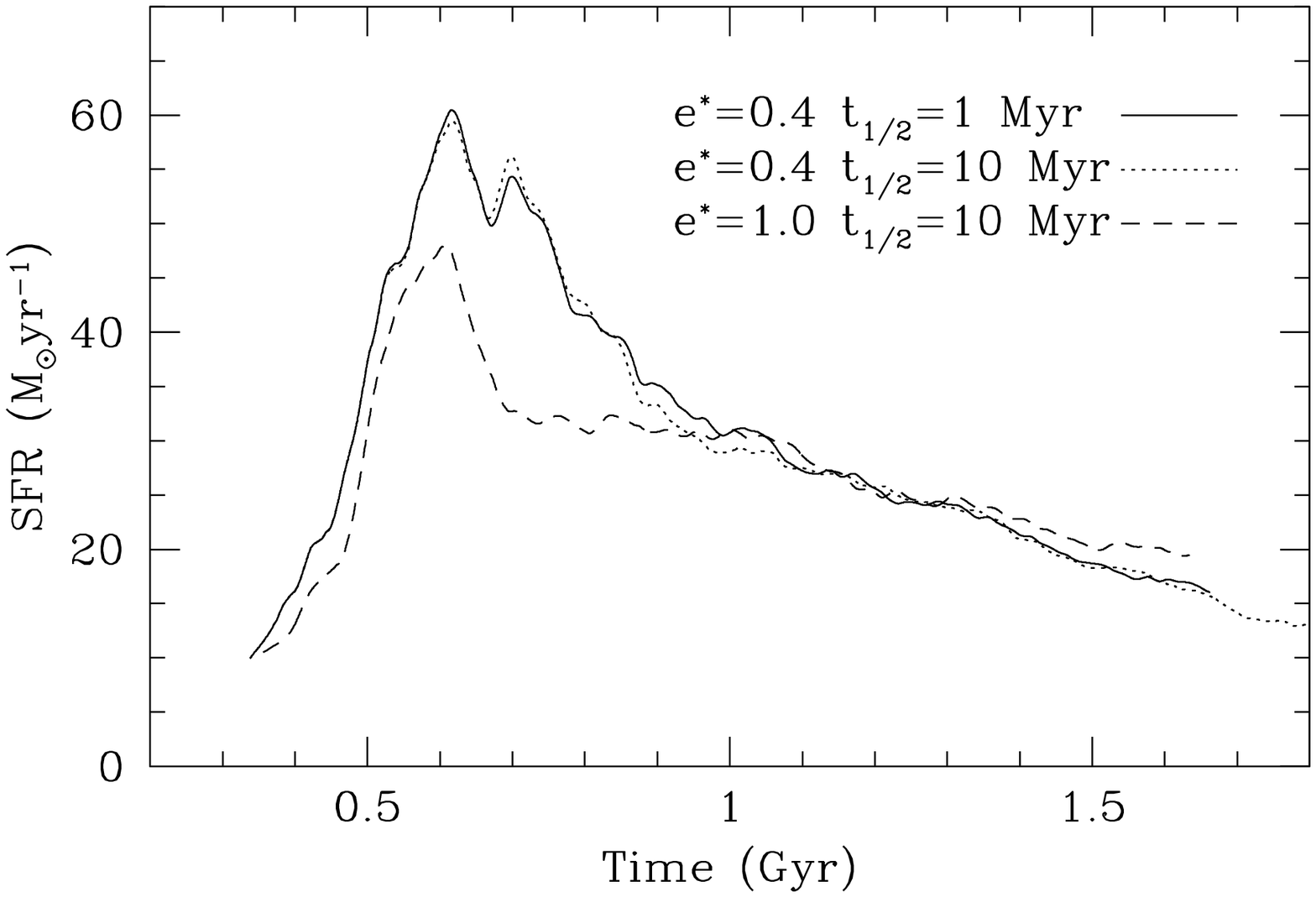}}

{\small {\footnotesize {\sc Fig.}~5.}---Dependence of
the SFR on the $t_{1/2}$ and $e^*$
parameters (for the ESa algorithm).  Comparing the runs with the same
$e^*$ parameter shows that varying $t_{1/2}$ from 1 to 10 has no
noticeable effect. Conversely, changing $e^*$ from 0.4 to 1.0 removes the
secondary and tertiary peaks in the SFR.
}
\bigskip

\noindent run produces enough heating to provide a significant amount of
expansion rendering $\nabla.{\bf v}\gg0$ for the first feedback region.
Note that since less of the gas is used at early times, the SFR at later
epochs is higher.

In summary, while the $c^*$ parameter clearly sets the SFR normalization,
it does not change the epoch of peak star formation. Further, the
$t_{1/2}$ parameter has little effect on the overall SFR due to the low
volume fraction of feedback regions in the evolved system.

\section{Application to isolated `realistic' models}\label{res}
This section reports the results of applying the algorithm to idealized
models of mature isolated galaxies. These models are created to fit the
observed parameters of such systems, \ie the rotation curve and disk scale
length. The characteristics of each model are discussed within the
sections devoted to them.  In this investigation the relative gas to
stellar fraction is low, compared to the primordial ratio, and thus star
formation is not as rapid as would be expected in the early stages of the
cosmological simulations (section~\ref{cosmomod}). Both models were
supplied by Dr.  Fabio Governato. We note that they both have a
sufficiently high particle number ($10^4$ SPH particles) to represent the
gas dynamic forces with reasonable accuracy (\cite{SM93,rob}). A summary
of the simulations is presented in table~2.

\subsection{Milky Way prototype}\label{mw}
The first prototype model is an idealized one of the Galaxy. It is desired
that the model should reproduce the measured SFR $\sim$ 1 M${}_\odot$
yr${}^{-1}$ and also the velocity dispersion in the disk. Evolved galaxies
have a lower relative gas content than protogalaxies. Further, because of
hierarchical clustering, they are significantly more massive. Hence
feedback is expected to have less effect on this model than on
protogalaxies formed in simulations of hierarchical merging. 

\subsubsection{Model Parameters}
The Milky Way prototype contains stars, gas and dark matter. The total
mass of each sector is $5\times10^{10}$ M${}_\odot$, \\

\begin{center}
\begin{tabular}{cclcc}
\hline\hline
Run & Simulation object${}^{\rm a}$ & feedback${}^{\rm b}$ &
$N_{step}^{\rm c}$ & $N_{SPH}$ \\
\hline
1001 & NGC 6503 & none & 3010 & 10240 \\
1002 & NGC 6503 & SPna & 3453 & 10240 \\
1003 & NGC 6503 & SPa  & 3986 & 10240 \\
1004 & NGC 6503 & TS   & 5345 & 10240 \\
1005 & NGC 6503 & ESna & 3453 & 10240 \\
1006 & NGC 6503 & ESa  & 4535 & 10240 \\
2001 & Milky Way & none & 387 & 10240 \\
2002 & Milky Way & SPna & 952 & 10240 \\
2003 & Milky Way & SPa  & 943 & 10240 \\
2004 & Milky Way & TS   & 723 & 10240 \\
2005 & Milky Way & ESna & 424 & 10240 \\
2006 & Milky Way & ESa  & 453 & 10240 \\
6001 & Cosmological & SPa & 4792 & 17165 \\
6002 & Cosmological & SPna  & 4710 & 17165 \\
6003 & Cosmological & ESa   & 4319 & 17165 \\
6004 & Cosmological & ESna & 4319 & 17165 \\
6005 & Cosmological & TSna  & 4335 & 17165 \\
6006 & Cosmological & TSa  & 4322 & 17165 \\
6007 & Cosmological & NF  & 4341 & 17165 \\
6008 & Cosmological & NSF  & 4314 & 17165 \\
6010 & Cosmological & TS  & 4342 & 17165 \\
\hline
\end{tabular}
\end{center}
\medskip
\noindent{\small \bf Table 2:}
{\small Summary of the main
simulations using
realistic models. ${}^{\rm a}$`Cosmological' refers to the object formed 
being derived from a cosmological simulation.
${}^{\rm b}$SPa=Single particle adiabatic period, SPna=single
particle no adiabatic period but adjusted cooling density, ESna=Total
energy smoothing with adjusted cooling density but no adiabatic period,
ESa=Total energy smoothing with adiabatic period, TS=Temperature smoothing
(normal cooling), NF=no feedback, NSF=no star formation.
${}^{\rm c}$For the cosmological simulations the number of
time-steps to $z=1.0$ is given. Since the isolated simulations were not  
run to the same final time, the average number of
time-steps per 100 Myr is shown.}
\bigskip

\noindent $9\times10^9$ M${}_\odot$, and $3\times10^{11}$ M${}_\odot$
respectively. 11980 particles were used to represent the stars, 10240 to
represent the gas and 10240 to represent the stars. The individual
particle masses were $4\times10^{6}$ M${}_\odot$, $9\times10^5$
M${}_\odot$, and $3\times10^{7}$ M${}_\odot$ respectively. The (stellar)
radial scale length was 3.5 kpc and the scale height 0.6 kpc. Density and
velocities were assigned using the method described in Hernquist (1993). 
The maximal radius of the dark matter halo was 85 kpc.  The artificial
viscosity was not shear-corrected in this simulation since the simulation
was integrated through only slightly more than two rotations and the
particle resolution in the object of interest is quite high. 

A comparatively large softening length of 0.5 kpc was used, rendering the
vertical structure of the disk poorly resolved. However, this is in line
with the softening lengths that are typically used in
cosmological simulations (of order 2 kpc). Shorter softening lengths allow
higher densities in the SPH, which in turn leads to higher SFRs. The
self-gravity requirement was again removed.

\subsubsection{Results}
In \fig~6, the gas particle distributions are shown for the
NF, SPa, ESa and TS runs. Of the versions not shown, ESna has a smooth
disk since the feedback regions do not persist as long and the SPna disk
resembles that from the SPa run (see section~\ref{profiles}).  TS produces
the most significant disturbance in the disk, which is to be expected
given that it injects more energy into the ISM than the other methods. For
TS feedback, 7\% of the disk gas had been ejected (falls outside an
arbitrary horizontal 6 kpc band) by t=323 Myr rising to 14\% by t=506 Myr.
Note that the amount of ejected gas is calculated relative to the total
gas in the simulation at the time of measurement. This is a decreasing
function of time, but is similar for all simulations since the SFRs differ
little.  Particle ejection velocities, $v_z$, were close to 300 \kms,
although some did achieve escape velocity ($\simeq 500$ \kms\ at solar
radius).  Hence, while TS can project particles out of the halo
(`blow-away') it preferentially leaves them bound in the halo
(`blow-out'). SP feedback (both SPa and SPna) has a similar evolution, but
only tends to eject the single heated particle during each feedback event,
thus leading to a lower mass-loss rate (1\% of the disk gas had been
ejected by t=323 Myr for both versions). Particles were often ejected with
$v_z > 600$ \kms, which is larger than the escape velocity, leading to a
proportionally stronger tendency for blow-away. ESa also ejects particles
from the disk (0.4\% ejected by t=323 Myr), although in general the
particles have lower velocities ( $v_z \simeq 200$ \kms) than the
particles ejected by either TS or SP. Hence, almost all of the ejected gas
remains bound to the system, \ie ES leads only to blow-out. ESna does not
eject particles since the feedback regions cool sufficiently fast (0.01\%
ejected by t=323 Myr).

\setcounter{figure}{5}
\begin{figure*}[t]
\epsfxsize=15.2cm
\centerline{\epsffile{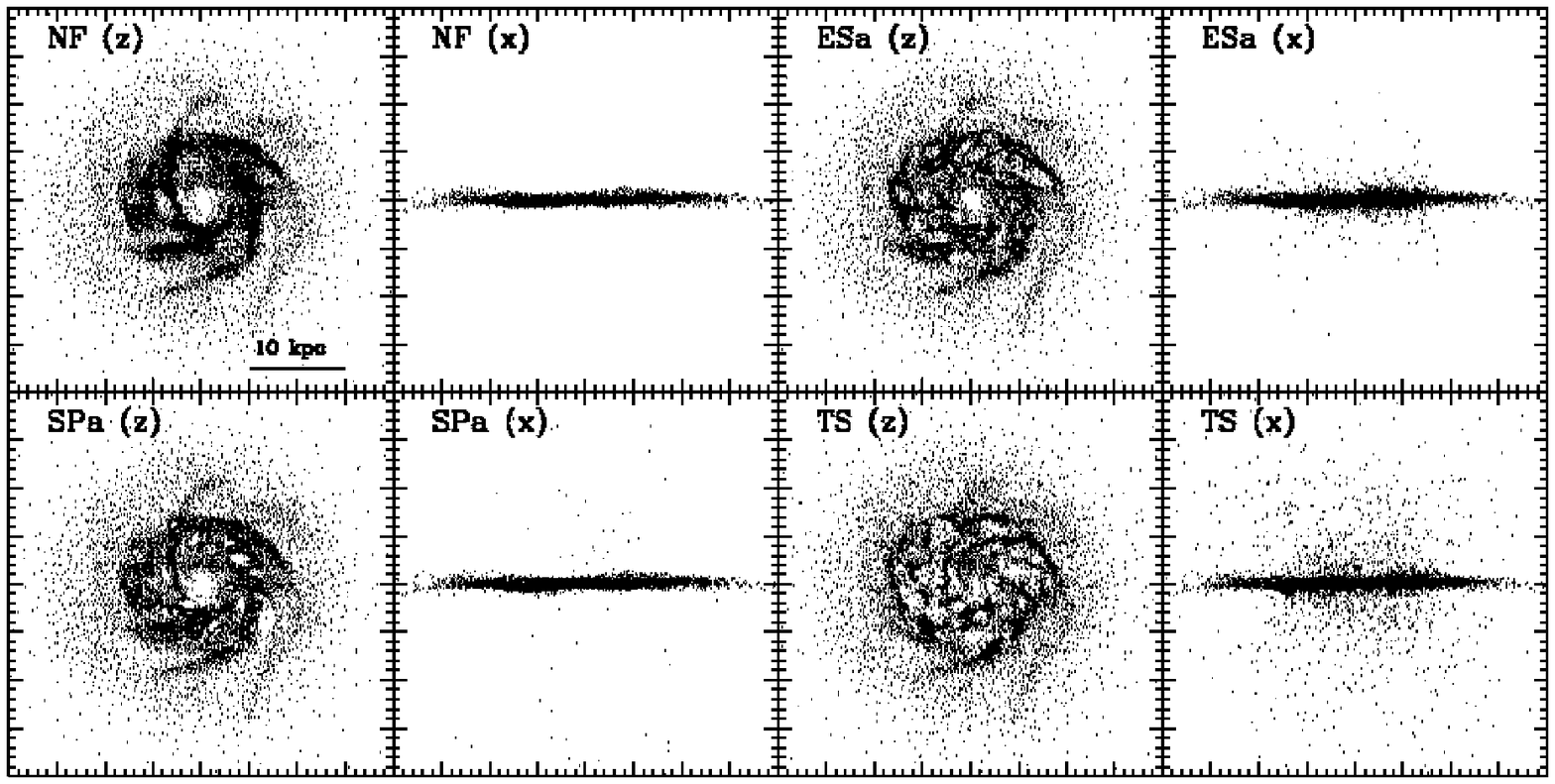}}
\caption{Morphology of Milky Way simulations at t=323 Myr. z- and
x-projections are shown to detail disk and gas halo structure. Both SPa
feedback and TS lead to significant ejection of matter from the disk. ESa
`inflates' the disk but does not eject as much matter as TS.}
\label{mwxzprofs}
\end{figure*}

The SFRs for three of the simulations (NF,TS and ESa) are plotted in
\fig~7. Most noticeable is the reduction in the SFR produced by TS and ESa
(TS is 35\% lower than no feedback at t=500 Myr, ESa is 10\% lower). This
is due to three factors; (a) the ejection of matter from the disk depletes
the cold gas available for star formation (see \fig~\ref{mwxzprofs}), (b)
smoothing feedback energy leads to spatially extended `puffy' feedback
regions in the disk, which in turn leads to a lower average density, and
hence lower SFR, (c) particles in the feedback regions will typically be
above the temperature threshold, which prevents star formation which
further reduces the SFR.  For single particle feedback the lower mass loss
rate leads to a higher SFR than for the TS or ESa runs. Of the versions
not plotted, ESna resembles the no feedback run (SFR approximately 3-4\%
lower on average), since most of the energy is rapidly radiated away. SPna
resembles the SPa since the feedback events produce very similar effects
(see section~\ref{profiles}).

To calculate radial profiles of the disks, an arbitrary plane of thickness
6 kpc was centered on the disk. This thickness ensured that the stellar
bulge was contained within the band.
Radial binning was then performed on this
data set using cylindrical bins.

In \fig~8, gas rotation curves are compared for the
simulations at t=323 Myr. To provide a fairly accurate depiction of the
rotation curve that would be measured the rotation curve was calculated by
radial averaging $|{\bf r}\times{\bf v}|/|{\bf r}|$ rather than by
calculating the circular velocity from $\sqrt{GM(<R)/R}$.  The main
drawback of this method is that in the core regions, where there are few
particles in the bins, the measurement can become `noisy'.  Clearly,
\fig~8 shows that there is little difference between schemes
(a maximum of 9\% at 4 scale lengths\-ignoring the \\

{\epsscale{0.95} \plotone{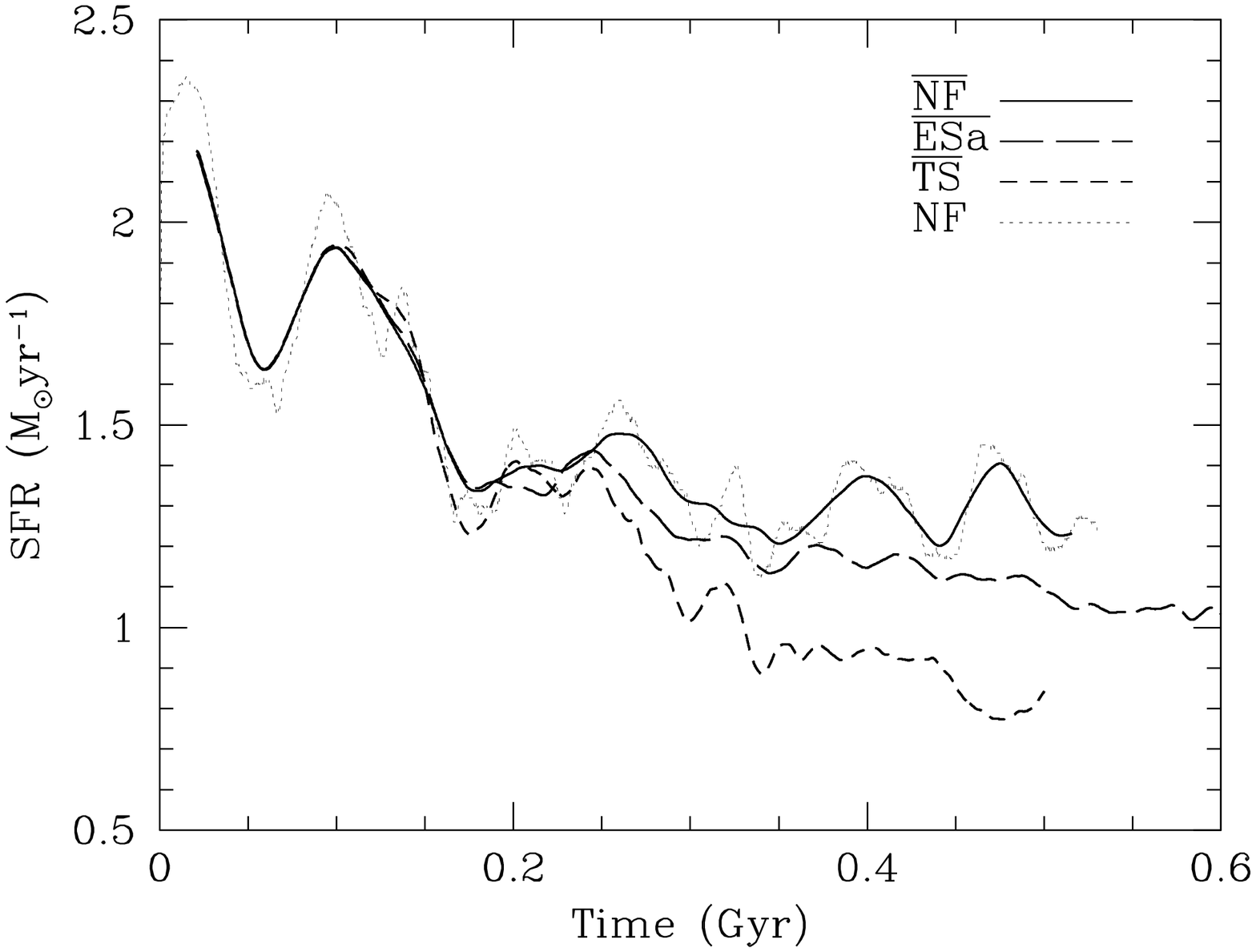}}

{\small {\footnotesize {\sc Fig.}~7.}---SFRs for the Milky Way prototype
(time-averaged
over 160 time-steps to show trend). The SP variants are not shown since
their evolution is similar to that of the ES version plotted. TS produces
a significant (35\% at t=500 Myr) reduction in the SFR as compared to no
feedback.  ES also reduces the SFR, but has a less significant effect
(10\% reduction at t=500 Myr versus no feedback) than TS.}
\bigskip

\noindent under-sampled central values). At large radii the curves match
precisely since there are no feedback events in the low-density outer
regions of the disk, except for the TS run where a feedback event has
ejected particles to the outer regions. Comparing to the initial rotation
curve (not shown), the disk has clearly relaxed, extending both in the
tail and toward the center. The curves were also examined at t=507 Myr
(for those simulations integrated that long) and similar results were
found with maximum differences being in the 10\% range. 

A more telling characteristic is the gas radial velocity dispersion. 
Unfortunately it is difficult to relate the measurements made here to
those of molecular clouds (\cite{MA95}), primarily because the mass scales
are significantly different. Nonetheless, it is interesting to compare
each of the separate feedback schemes. In \fig~8 the radial
velocity dispersion, $\sigma_r$, is plotted for three of the simulations
at t=323 Myr (again only considering matter within the 6 kpc band).
Temperature smoothing produces a large amount of dispersion due to the
excessive energy input (interior to 8 scale lengths it varies between
being 40\% to 300\% higher than other values). Note that there is a direct
correlation between a higher velocity dispersion and a lowered measured
rotation curve. This is asymmetric drift: feedback events produce velocity
dispersion which in turn increases the relative drift speed, $v_a$,
between the gas and the local circular velocity (\cite{BT87}). The
remaining algorithms (SPa, SPna, ESa, ESna) exhibit similar velocity
dispersions. Thus, for all but the TS algorithm, the bulk dynamics remain
the most important factor in determining the velocity dispersion. Although
the velocity dispersion presented here is not directly compatible with
that of the value for local molecular clouds, it is interesting to note
that measurements for the Milky Way suggest $\sigma_{v_{cloud}}= 9\pm1$
\kms (\cite{MA95}). 

Due to the finite size of the computational grid, the code was unable to
follow all of the simulations to the desired final epoch (500 Myr). This
limitation was most noticeable in the SP simulations where ejected gas
particles reached the edge of the computational domain within 320 Myr. It
is possible to remove these particles from the simulation since they are
comparatively unimportant to the remainder of the simulation.  However
maintaining an accurate integration was considered to be more important. 

{\epsscale{0.95}
\plotone{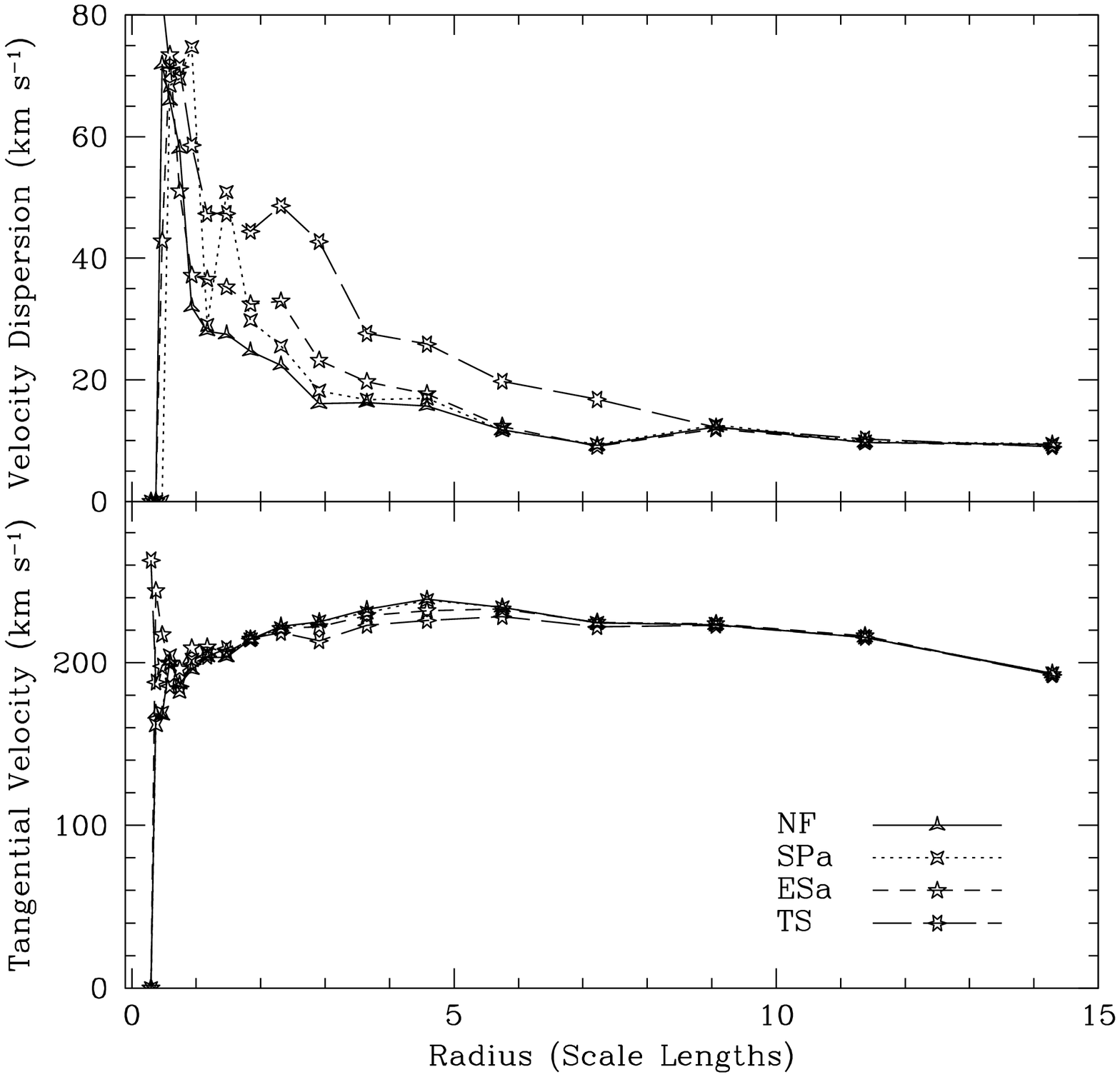}}

{\small {\footnotesize {\sc Fig.}~8.}---Rotation curves and radial
velocity dispersions
for the Milky Way prototype at t=323 Myr, for the NF, SPa, ESa and TS  
runs. There is only a marginal difference between rotation curves because 
the radial averaging smooths out the effect of inhomogeneous feedback
regions. The TS gas exhibits a 8\% reduction in the rotation curve at
a radius of 4.5 scale lengths due to asymmetric drift. The ESa rotation
curve is reduced by only 4\% at this radius. The TS algorithm
exhibits significantly higher velocity dispersion than any of the other  
variants. This is attributable to the large energy input driving winds  
that strongly affect the disk structure (visible in \fig~\ref{mwxzprofs}
as large holes in the disk). All the other algorithms differ only very
marginally.}
\bigskip

\subsubsection{Summary}
 Temperature smoothing (TS) is clearly the most violent feedback method,
producing an SFR lower than the other algorithms and also tending to
evaporate the disk. Given the comparatively large escape velocity of the
Milky Way (lower limit of 500 km s${}^{-1}$ at the solar radius), this is
an unrealistic model because such dramatic losses are expected only in
dwarf systems. Of the remaining algorithms, the single particle (SP)
versions produce reasonable physical characteristics, but have the
disadvantage of requiring a large number of time-steps. The energy
smoothing variant with an adiabatic period (ESa) appears to be the best
compromise in these simulations. It does not require an excessive number
of time-steps while the disk morphology and evolution are within
reasonable bounds: there is no excessive blow-out or blow-away. 

\subsection{Dwarf prototype}
The second model is an idealized version of NGC 6503. Dwarf systems are
expected to be more sensitive to feedback due to their low mass, and
consequently lower escape velocity.  In simulations, the over-cooling
problem suggests that to form a large disk system from the merger of small
dwarfs, the dwarf systems must have significant extent (ideally similar to
that for an adiabatic system, \cite{WE98}). Feedback is currently believed
to be the best method for achieving this. Given that for NGC 6503
$v_c\simeq 110$ \kms\, a lower bound on the escape velocity of the system
is 155 \kms. 

Detailed numerical studies of NGC 6503 have been conducted by Bottema and
Gerritsen (1997)  and Gerritsen (1997). The motivation in this
investigation is different to the previous ones which attempted to explain
the observed dynamics of NGC 6503.  In contrast, this investigation
attempts to determine bulk properties at comparatively low resolution, in
accordance with that found in cosmological simulations. 

\subsubsection{Model Parameters}
As for the Milky Way prototype, this model contains stars, gas and dark
matter, with the total masses of $3\times10^{9}$ M${}_\odot$,
$1\times10^9$ M${}_\odot$, and $5\times10^{10}$ M${}_\odot$ respectively.
10240 particles were used in each sector, yielding individual particle
masses of $3\times10^{5}$ M${}_\odot$, $1\times10^5$ M${}_\odot$, and
$5\times10^{6}$ M${}_\odot$, respectively. The radial scale length of the
simulation was 1.16 kpc, and the scale height 0.1 kpc.  Gas density and
particle velocities were assigned in the same fashion as the Milky Way
prototype. The artificial viscosity was not shear-corrected for the same
reasons as the Milky Way protoype.

Six simulations were run, each using a different method of feedback
(including no feedback as the control experiment). The method used in each
simulation and the number of time-steps per 100 Myr are summarized in
table~2.

\subsubsection{Results}
As in the Milky Way simulations, the SP feedback models produced
significant blow-away at the outset of the simulation and particles
ejected from the disk rapidly escaped from the halo.  Consequently, the
evolution of these systems had to be halted at very early times (close to
200 Myr).  Of the remaining algorithms, only TS was not integrated to at
least 500 Myr. Thus, the following analysis concentrates on the ES
variants. 

\setcounter{figure}{8}
\begin{figure*}[t]
\epsfxsize=11cm
\centerline{\epsffile{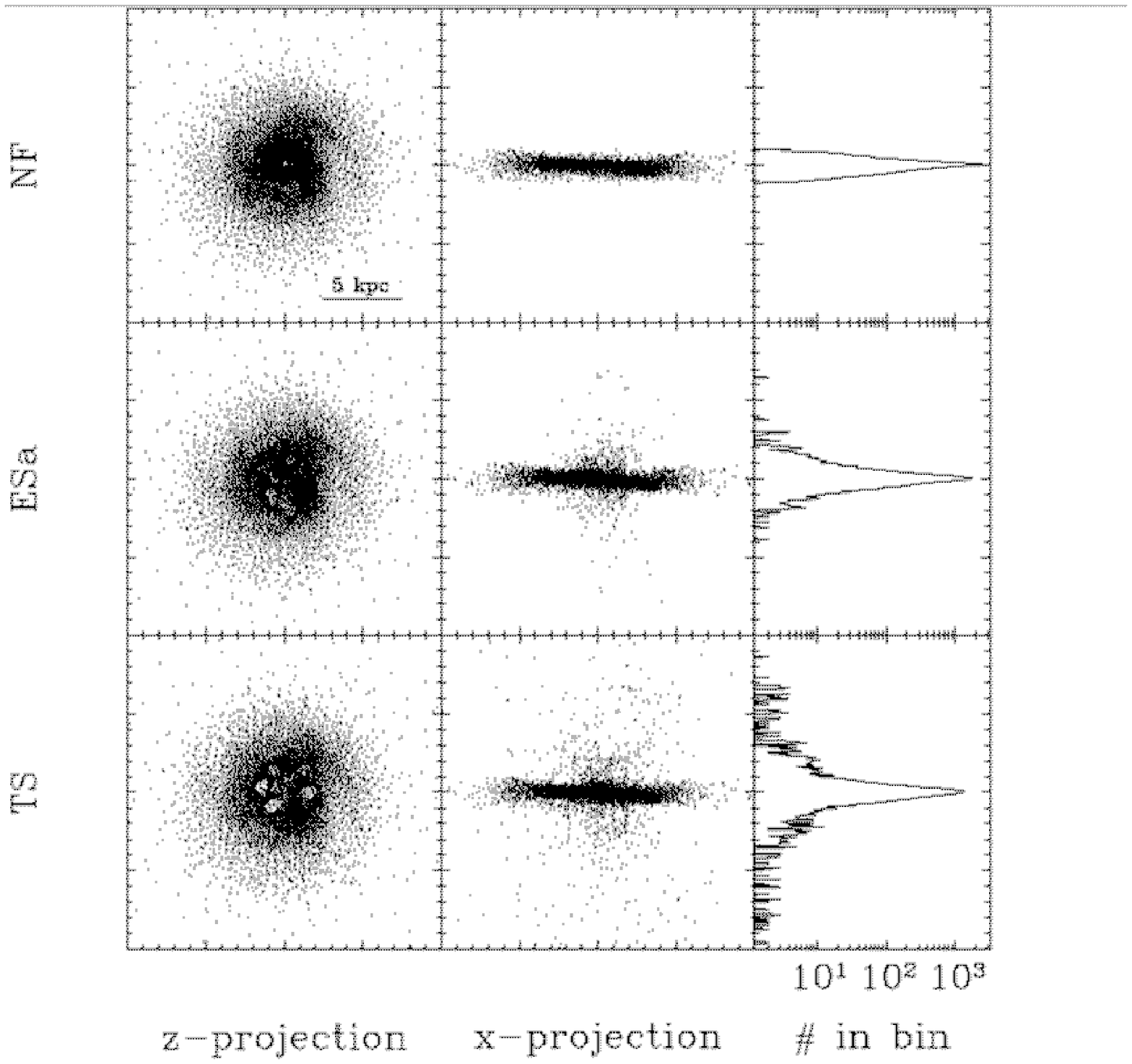}}
\caption{Selected
morphologies for the NGC 6503 model. The left hand
hand column shows the z-projection of  
gas particles in the NGC 6503
simulation, the central column shows the x-projection and the right column
displays the z-coordinate of the gas particles in the NGC 6503 simulation
in bins of width 0.150 kpc ($h_{min}/2$).
A significant amount of
relaxation is visible in the no feedback run, with the disk fattening
to a
width of $\simeq 600 pc$. Feedback in the TS and ESa runs ejects
particles
into the halo and populates the tails of the z-distribution of particles.}   
\label{zdist}
\end{figure*}

\Fig~\ref{zdist} shows the distribution of gas particles in the x-
and z-projections and also the z-distribution measured vertically in
bins,
at t=230 Myr (note that although sufficient time has elapsed for feedback
events to occur the disks still exhibit virtually identical rotation
curves).  Due to the comparatively long softening (300 kpc) length used in
the simulation, there is a significant amount of relaxation from the
initially `thin' distribution, which is approximately 250 pc wide. Since
in the simulation code the SPH resolution is at least twice the
gravitational
softening length,
the disk was expected to fatten to 600 pc.  \Fig~\ref{zdist}
shows that this is observed (in the run with no feedback). Once feedback
is included, and matter is ejected from the disk, the z-distributions
develop populated tails due to particles orbiting high in the potential
well. The most severe example of this being the TS
variant, which rapidly ejects particles leading to significant mass loss
from the disk (both blow-away and blow-out occur). The TS
algorithm produces extremely large bubbles in the disk. One
bubble had a radius of almost 0.7 kpc, which is 60\% of the disk scale
length, while for the Milky Way prototype larger bubbles had a radius of
about 0.8 kpc, approximately 40\% of the scale length.
The absolute size of these bubbles relative
to the disk is set by the SPH smoothing scale and hence should not be
overinterpreted. However, the
 comparison of the dwarf versus the Milky Way model is valid since the
particle resolution is approximately the same for both models. A
comparison of the gas distributions for the ESa runs (dwarf {\em vs.} 
Milky Way) 
at 500 Myr shows that while in the dwarf system the gas density puffs up
beyond
the stellar component, it does not do this for the Milky Way prototype. 
As in the Milky Way runs, ES preferentially leads to
blow-out, although by 500 Myr some particles were close to escaping the
halo.  These results show that feedback has a more significant effect on
the
dwarf system.

To calculate radial profiles of the disks a plane of thickness 2 kpc was
used, centered on the disk. As for the Milky Way prototype, the thickness
was chosen to ensure that the stellar content was included within the
band. The data were again binned using concentric cylinders.  At t=580 Myr
the gas rotation curves for ESna, ESa and no feedback remain very similar
(\fig~10). Both of the curves exhibit asymmetric drift
relative to the run with no feedback. The maximum difference (external to
a radius of one scale length) occurs at 1.6 scale lengths where the
adiabatic variant has a rotation curve that is 10\% lower than the no
feedback run.  At this radius the non-adiabatic run is only 3\% lower than
the run with no feedback. The outer edges of the distributions remain
identical due to there being no feedback events in the low-density gas.

The gas radial velocity dispersion plot (\fig~10) shows that
for this system ESa clearly introduces more dispersion (30\% higher at a
radius of 1.6 scale lengths). This is as expected: the combination of a
lower escape velocity and comparatively long persistence of the feedback
regions in the adiabatic variant allow the gas to escape to higher regions
of the potential well. Additionally, bubble expansion in the plane of the
disk persists for longer in the adiabatic variant.  Notably, the run with
no feedback shows an increasing velocity dispersion with radius. This can
be attributed to the large softening length used, which in turn causes the
SPH to smooth over a very large number of particles in the central regions
(in excess of $10N_{smooth}$). Thus, in this region the gas distribution
is dynamically cold. 

\Fig~11 shows the time-averaged SFRs for the three runs.  By
t=580 Myr the non-adiabatic run had ejected 1\% of its mass from the disk
(relative to the remaining gas), whilst the adiabatic run had ejected 2\%. 
Examination of the raw data shows that the strongest bursting is actually
found in the no feedback run. The feedback in the other two runs keeps the
disk more stable against local collapse. Of the data not plotted, TS was
similar to the ESa run, and had an SFR approximately 5\% lower. By t=240
Myr, 6\% of the disk had been evaporated. Neither of the SP runs was
integrated far enough to detect \\

{\epsscale{0.95}
\plotone{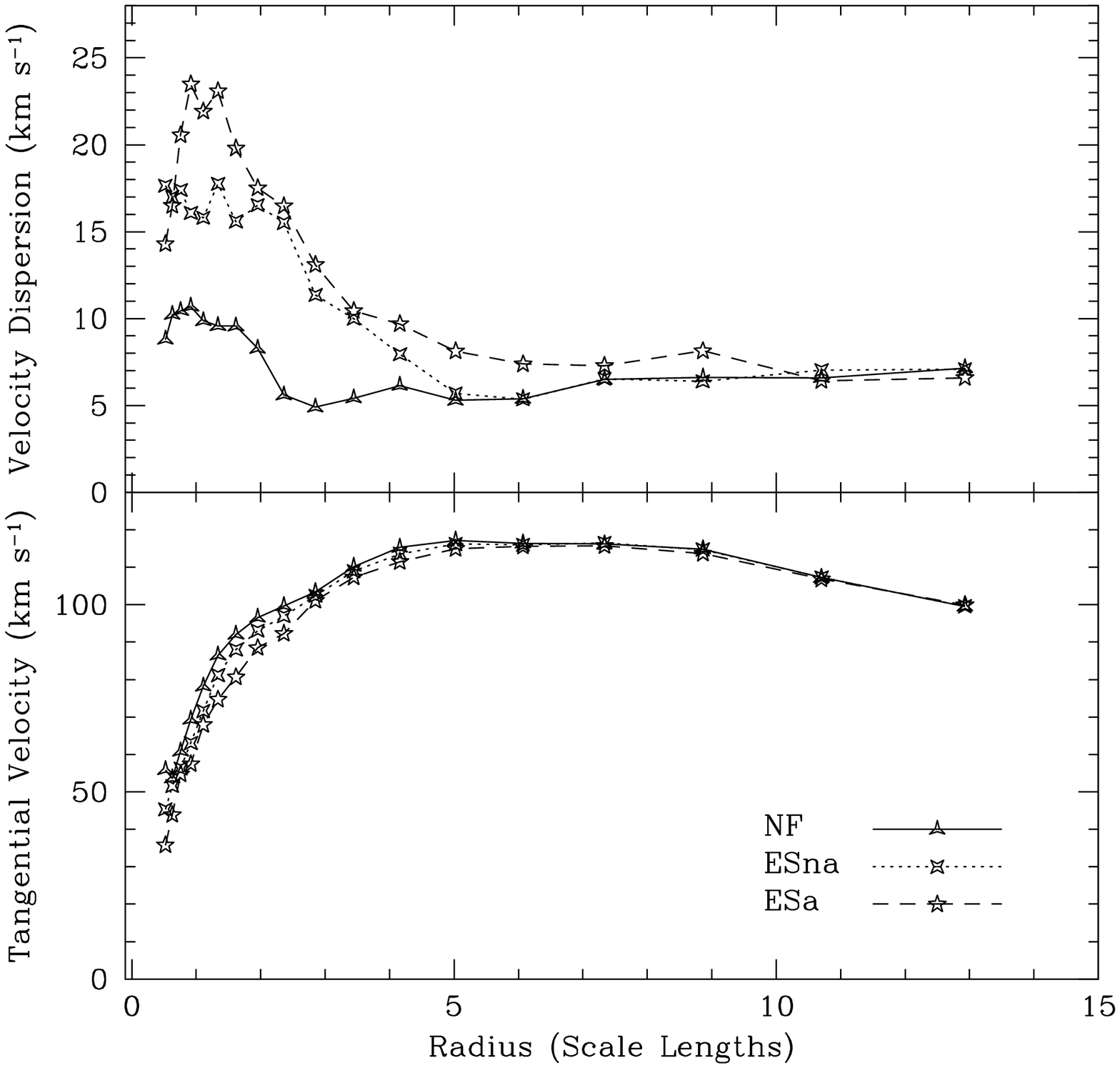}}

{\small {\footnotesize {\sc Fig.}~10.}---Comparison of
rotation curves and radial velocity
dispersions for the NGC 6503 dwarf simulation at t=580 Myr. There is 
little difference among all rotation curves, except in the nuclear   
region where feedback is more prevalent. Asymmetric drift is visible in  
the gas, with the ESa run having a lower rotation curve due to its higher
velocity dispersion.}
\bigskip

{\epsscale{0.95}
\plotone{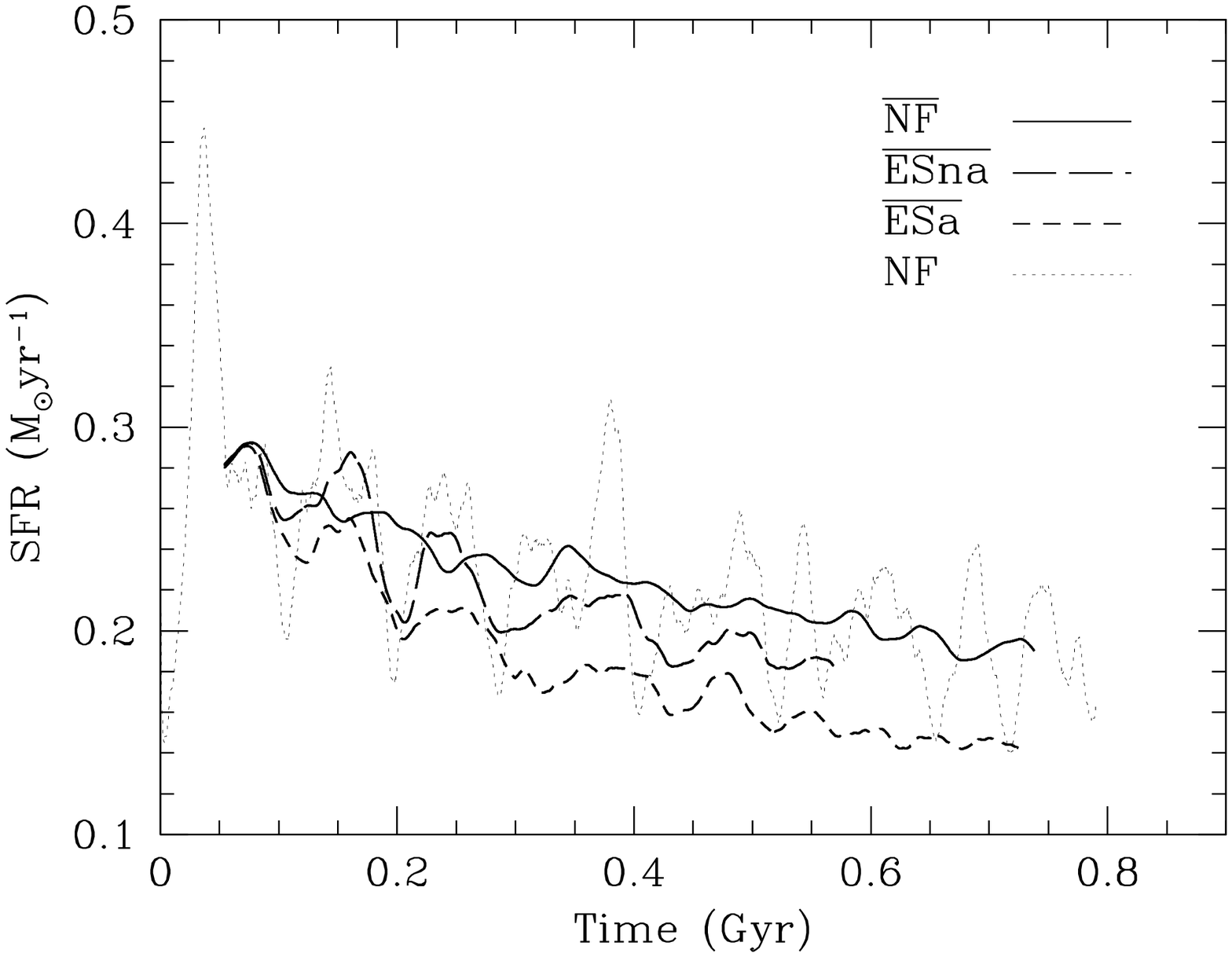}}

{\small {\footnotesize {\sc Fig.}~11.}---SFRs for the dwarf prototype. The
ES and no
feedback runs are shown since the remainder were not integrated for a
sufficient time for conclusions to be drawn. The NF, ESa, and ESna runs
were time-averaged over 320 time-steps to elucidate the trend in the SFR
and this is indicated by the bar symbol in the legend.  Clearly the ESa
run produces the lowest SFR, being approximately 20\% lower than the
non-adiabatic run.}
\bigskip

\noindent significant trends. 

\subsubsection{Summary}
Although it was not possible to integrate all the models to the desired
final epoch, it was still clearly demonstrated that feedback does have a
more significant impact on the dwarf system. This is evident both in the
morphology (larger relative bubbles as compared the Milky Way prototype)
and radial characteristics (the radial velocity dispersion is far higher
relative to the no feedback run). This increased sensitivity also allows
differentiation between the adiabatic and non-adiabatic methods, which was
at times difficult in the Milky Way prototype. Whether these conclusions
can carry over to cosmological simulation is addressed in the next
section.

\section{Cosmological simulations}\label{cosmomod}
As discussed in the introduction, there are a number of problems that
plague cosmological simulations of galaxy formation. This section examines
the conjecture that, following an initial burst of star formation,
feedback should be able to (1) prevent the overcooling catastrophe by
suppressing {\em massive} early star formation and (2)  prevent the
angular momentum catastrophe, thereby allowing the formation of disks with
specific angular momenta in agreement with observations. We study all of
the feedback algorithms analysed in the previous sections and also include
two new versions derived from combinations of the previously studied
algorithms. Note that the SPH resolution ($\sim2\times10^3$ particles) in
the galaxies formed in the following simulations is insufficient to
resolve shocks adequately.  The purpose of the simulations is to explore
the parameter space and not make precise predictions about resulting
galaxies. A higher resolution simulation, meeting the accuracy criteria
outlined in Steinmetz \& Muller (1993), will be presented in a subsequent
paper. 

\subsection{Initial conditions}
In the SPH method, shocks are captured using an artificial viscosity. The
artificial viscosity is turned on or off by the value of the ${\bf r}.{\bf
v}$ product between each pair of particles. The angular part of this
product takes a maximal value if {\bf r} and {\bf v} are aligned, as is
the case in collapse along a Cartesian grid. Collapse along a direction
not aligned along the grid leads to scatter in the ${\bf r}.{\bf v}$ dot
product, and hence less shocking. Consequently, we believe that it is
advantageous to use a set of initial conditions must be used that contains
no preferred direction. `Glass-like' initial conditions are used in this
study. 

Given a hierarchical clustering scenario the first objects to form will
have hundreds (at most) particles in them. Hence it is only necessary to
create initial conditions which have no preferred direction on scales of
the order 500 particles. The merging of the first generation progenitors
occurs over scales significantly larger than a grid spacing, thus removing
concerns about a preferred collapse direction for these objects.  It thus
makes sense to create a small periodic glass with very low noise and then
tile this within the simulation box.

To create the glass `tile', 512 particles are placed in a periodic
box. These particles are forced to be anti-correlated by not allowing any
two particles to be closer than $0.9\times N^{1/3}$, which is 0.9 times
the average inter-particle spacing. This initial condition has a very low
noise level. The noise level is further reduced by evolving the glass in a
(periodic)  gravity-only simulation with the sign of the velocity update
reversed. With this modification, the particles repel one another,
and eventually relax \\

{\epsscale{0.88}
\plotone{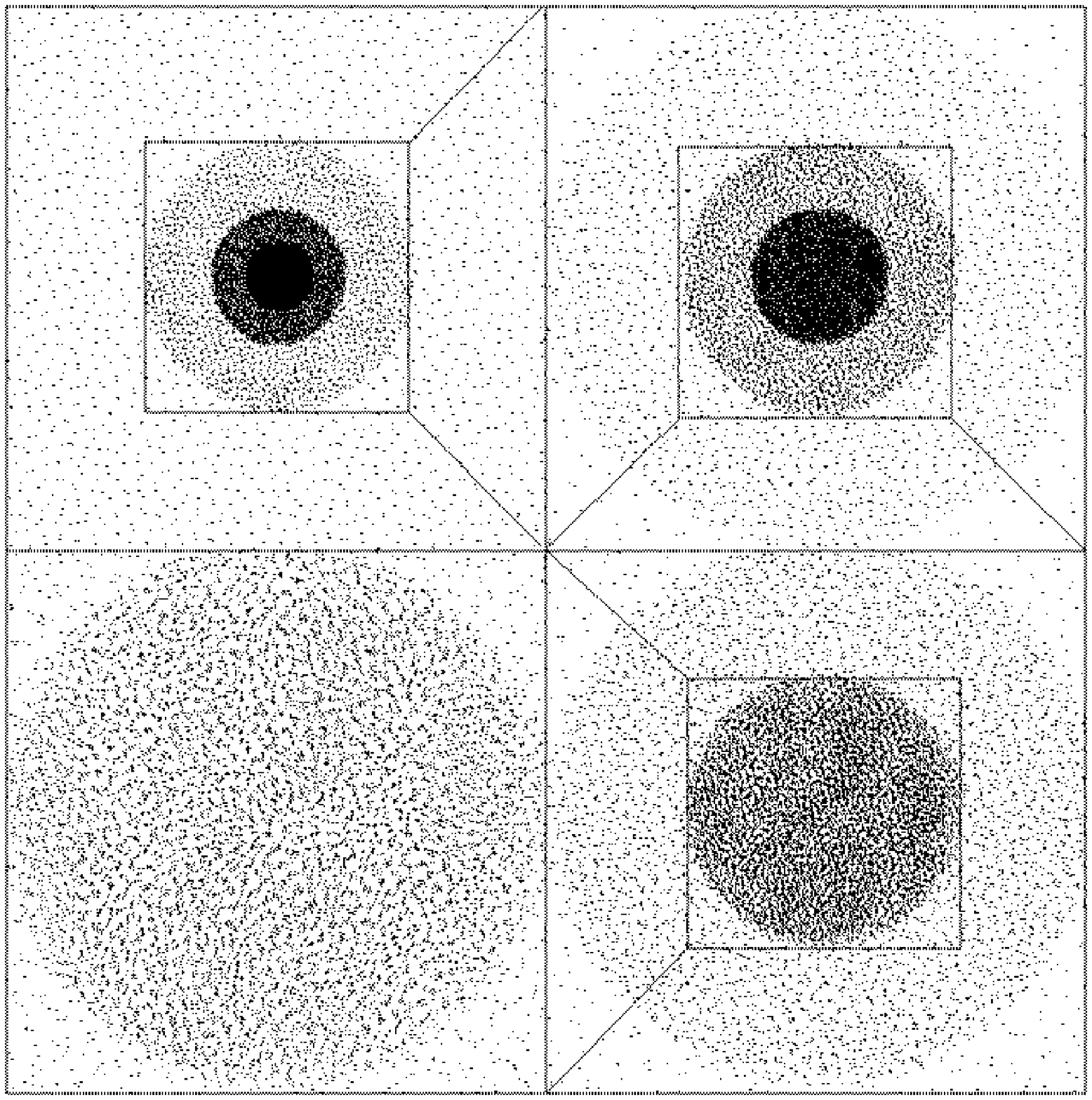}}

{\small {\footnotesize {\sc Fig.}~12.}---Layering of cosmological initial
conditions.  Starting clockwise from the upper left panel, the top level
configuration of $32^3$ particles is repeatedly cut and shrunk to create a
hierarchy four levels deep. Gas particles are included in the central
region only.} \bigskip

\noindent to a state in which the (repulsive) potential energy is reduced
to a minimum. 

Once the tile is fully evolved, it is replicated a number of times to form
the main simulation cube. This configuration does not have any noise on
scales larger than the size of tile and thus constitutes an excellent
initial configuration. Because long-range tidal forces have a significant
effect on the evolution of galaxies (\cite{KP95}) they must be included in
simulations. Unfortunately, a fixed resolution periodic box with equal
number of dark matter and gas particles would require a prohibitively
large number of particles. Hence we used the multiple mass technique
(\cite{port85}) to overcome this problem.  The hierarchical layers are
constructed by successively cutting out a region of the simulation cube
and replacing it with a copy of the top-level `grid' cut and shrunk to the
appropriate size.  The first layer, for example, is constructed by
removing a sphere of radius $1/4$ the box size and then filling that
region with a sphere cut from the main simulation cube and shrunk to size.
The next layer is formed by cutting a sphere of radius $1/8$ the box and
replacing this with a similarly cut and shrunk sphere from the main
simulation cube.

Unfortunately this process does introduce some noise at the boundary of
each region. It was thus assured that in the highest resolution region,
objects of interest form sufficiently far away from the boundary with the
next region. The layering process continues through four layers. To
maintain mass resolution, the particles in each layer have mass $1/8$ that
of the previous layer. Thus the mass resolution of this region is 512
times higher than the lowest resolution region, and the spatial resolution
is eight times higher.  \Fig~12 shows the layering in detail.
If a grid of $32^3$ particles is initially used to perform the layering,
the resulting system has 77,813 dark matter particles and 17,165 SPH
particles. In the high resolution region the effective resolution is
$2\times 256^3$. 

Assigning the perturbations associated with the initial power spectrum is
more difficult for multiple mass simulations. The particle Nyquist
frequency is not constant across the simulation. If the box is loaded with
modes up to Nyquist frequency of the highest resolution region, then
aliasing of the extra modes will occur in the low
resolution region. Hence, to prevent aliasing, the lower resolution
regions must have their displacements evaluated from a force grid that is
calculated using only modes up to the local particle Nyquist. Thus all of
the box modes are calculated, and then modes are progressively removed by
applying a top-hat filter in k-space.

\subsection{Simulation Parameters}
To assign adiabatic gravitational perturbations, the linear CDM
power spectrum of Bond and Efstathiou (1984) was utilized,
\begin{equation}
P(k)={Aq \over [1+(23.1q+(11.4q)^{3/2}+(6.5q)^2)^{5/4}]^{8/5}},
\end{equation}
where,
\begin{equation}
q=k/(\Gamma h).
\end{equation}
Given a baryon fraction of 10\% the shape parameter, $\Gamma$, was 
calculated
from Vianna and Liddle (1996) yielding $\Gamma\simeq0.41$. The Hubble
constant was set at 50 km s${}^{-1}$ Mpc${}^{-1}$, yielding in
$h=0.5$ in the standard H${}_0$=$100h$ km s${}^{-1}$ Mpc${}^{-1}$ units. 
The normalization constant, $A$, was chosen so as to reproduce the
number density of rich clusters as observed today, which is
given by the rms mass variance $\sigma_8=0.6$
(\cite{ecf}). The initial redshift was
$z=67$ and the box size 50 Mpc (all length scales are quoted in
real units). 

\setcounter{figure}{12}
\begin{figure*}[t]
\epsfxsize=16cm
\centerline{\epsffile{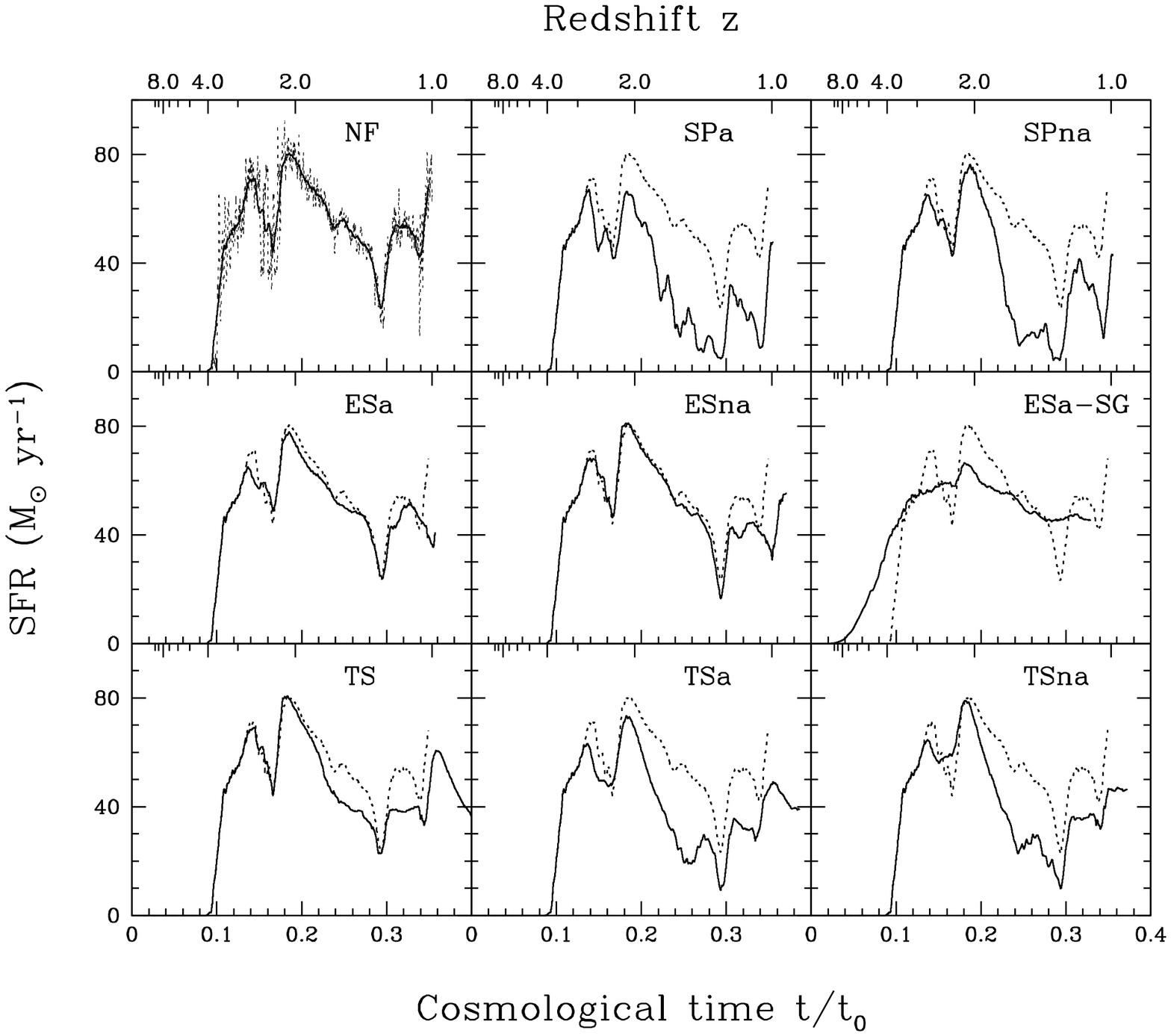}}   
\caption[Integrated SFRs for the entire high resolution region for the  
cosmological simulations.]{
SFRs for the cosmological simulations. The SFR shown is integrated over
the entire gas sector of the simulation ($8\times 10^{11}$ \msol). A 160
time-step average is used to smooth the data and more clearly elucidate 
trends. The effect of the smoothing is demonstrated in
the no feedback panel. For comparison, the smoothed NF SFR is
plotted as a dotted line in the remainder of the panels.}
\label{cosmo.SFR}
\end{figure*}

To ensure the collisionless nature of the dark matter does not become
contaminated by two-body forces, the two-body relaxation should be longer
than the Hubble time. Thomas and Couchman (1992) show that for
a uniform distribution of particles (of mass m, and softening $\epsilon$)
within a spherical volume of radius $R$ and with a velocity dispersion
$v^2 \sim \gamma G m N / R$ ($\gamma$ is a constant dependent upon the
characteristics of the velocity distribution) the two-body relaxation time
is
\begin{equation}\label{trelax}
t_{r}\simeq {\gamma^{3/2} N^{1/2} \over 6 \ln(R/\epsilon) } \left( {R
\over \epsilon} \right)^{3/2} t_2,
\end{equation}
where $t_2=(\epsilon^3 / Gm)^{1/2}$ is the minimum time-scale for
interactions of particles under gravity, with an effective impact
parameter $\epsilon$.
Utilising the velocity dispersion for an isothermal sphere and
modifying the $t_2$ parameterization of TC92, equation
\ref{trelax} may be rearranged to give
\begin{equation}
 {t_{r} \over t_0}
\simeq 0.02 { (R/\epsilon)^{3/2} \over ln(R/\epsilon) } N^{1/2} 
\left( {\epsilon \over 4 {\rm \; kpc} } {50 {\rm \; Mpc} \over box}
\right)^{3/2} \left( {N_p^{eff} \over 256^3 \Omega_p} \right)^{1/2}.
\end{equation}  
Ideally the ratio $t_{r}/ t_0$ should be greater than one, although since
the simulation evolves to $z=1$, values of $0.5$ should be
acceptable. 
The formula is minimized when $R={\rm e}^{1/2} \epsilon$ (\ie close to the
softening), and at this radius it is found that $t_r / t_0 \simeq 0.08
N^{1/2}$. This implies that within the softening volume, $N>10$ is
neccesary to
avoid two-body relaxation, although if the simulations were integrated to
$z=0$, the criterion would be $N>30$.

The effective resolution of the high resolution region (which is 6.25 Mpc
in diameter) is $256^3$, which yields a mass resolution of $4.6\times10^8$
\msol\ in the dark matter, $5.1\times 10^7$ \msol\ in the gas (reducing to
$2.6\times10^7$ \msol\ after the creation of the first star particle) and
$2.6\times10^7$ \msol\ in the star particles.  Clearly the mass
resolution
remains low, with a $10^{11}$ \msol\ galaxy (in baryons) being
represented by approximately 4,000 gas and star particles, assuming an
equal division of both. Nonetheless, this resolution is sufficient to
give a reasonable indication of the performance of different algorithms in
a cosmological environment. The total baryonic mass in the high resolution
region is approximately $8\times10^{11}$ \msol. This is a
consequence of attempting to keep the boundary of the second mass
hierarchy a sufficiently long way from the object of interest, and
 choosing a sufficiently large box size to provide a
reasonable representation of tidal forces. Since the simulated disk will
was evolved over 3 Gyr, and the simulation had
comparatively low resolution, shear-correction was applied to the
artificial viscosity.

The first simulation conducted was a low resolution $128^3$ dark matter
simulation using the parameters given. From this simulation, candidate
halos for re-simulation were extracted at a redshift of $z=1$. Due to
wall-clock limits on simulation time, it was decided that $z=1$ was the
most appropriate time to stop the simulation. Whilst the dark matter run
could have be continued to $z=0$, this is prohibitively expensive for the
high resolution hydrodynamic runs, since it requires close to 15,000
time-steps. The chosen halo had a mass of $2.7\times10^{12}$ \msol\ and is
thus comparatively large. Re-simulation showed that it corresponds to the
halo of a merger event of two galaxies with a combined bayonic mass of
$1.8\times10^{11}$ \msol. 

\setcounter{table}{2}
\begin{table*}[t]
\centering
\begin{tabular}{cclcccccc}
\hline\hline
Run & feedback${}^{\rm a}$  & $R_{200}$  &
$\Sigma m_*$ &
$|L_{gc}|/|L_{dm}|$ &
$M_{\rm disk}$ &
$R_{disk}^{\rm b}$ &
$R_{inner}^{\rm c}$ &
$R_{outer}$ \\
${}$ & ${}$ & (kpc) & ($10^{10}$ \msol) & & ($10^{11}$ \msol) & (kpc) &
(kpc) & (kpc) \\
\hline
6001 & SPa & 187 & $7.98$ & 0.25 & $1.07$ &
8.8  & 1.0  & 34  \\
6002 & SPna & 187 & $8.44$ & 0.24 & $1.14$ &
7.6  & 0.7  & 30  \\
6003 & ESa & 188  & $9.51$ & 0.09 & $1.38$ &
8.0  & 0.6  & 17  \\ 
6004 & ESna & 188 & $9.38$ & 0.18 & $1.42$ &
11.3${}^\dagger$  & 0.7  & 24  \\
6005 & TSna & 189 & $7.45$ & 0.19 & $1.35$ &
9.3  & 0.8  & 27  \\
6006 & TSa & 188 & $7.25$ & 0.27 & $1.27$ &
9.3  & 0.9  & 24  \\
6007 & NF & 189 & $9.67$ & 0.16 & $1.57$ &
9.5  & 0.9  & 34  \\
6008 & NSF & 188 & N/A & 0.14 & $1.29$ &
9.3  & 0.6  & 28  \\
6010 & TS & 188 & $8.69$ & 0.19 & $1.38$ &
9.7  & 0.6  & 21  \\
 &  &  &  &  &  &  &   &   \\
6014 & ESa-SG  & 189 & $9.78$ & 0.18 & $1.25$ &
9.6  &  1.1  & 45  \\
6015 & ESa-2c*  & 189 & $10.9$ & 0.15 & $1.24$ &
9.1   & 0.9  & 50  \\
6016 &TSa-SG-2c*& 188 & $6.33$ & 0.33 & $1.06$ &
10.3  & 1.3  & 39  \\
6017 &ESa-SG-2c*& 189 & $11.6$ & 0.08 & $1.35$ &
9.2   &  0.8 & 36  \\
6018 & ESa-nav  & 188 & $3.83$ & 0.23 & $1.34$ &
8.2   &  0.9 & 36  \\
\hline
\end{tabular}
\caption[Summary of the properties of cosmological simulations at
z=1.09.]{Summary of the properties of cosmological simulations at z=1.09.
${}^{\rm a}$SPa=Single particle adiabatic period, SPna=single
particle no adiabatic period but adjusted cooling density, ESna=Total
energy smoothing with adjusted cooling density but no adiabatic period,
ESa=Total energy smoothing with adiabatic period, TS=Temperature smoothing
(normal cooling).   
${}^{\rm b}$The disk radius was evaluated by finding the radius at
which the baryon surface density fell below $2\times10^{13}$ \msol\
Mpc${}^{-2}$. This value was established by visually judging the edge
of the NSF disk and then reading off the surface density at this boundary.
${}^{\rm c}$The inner values are distorted to shorter scale
lengths by the presence, or lack of, a central core in the disk. The outer
fits are strongly affected by companion systems inside $r_{200}$ and 
strong emphasis should not be placed on these results.
${}^{\dagger}$This value is anomalously high due to a minor
merger
during which the stellar content of the merging dwarf orbits out at $>10$
kpc.}
\label{tbl-3}
\end{table*}

\subsection{System evolution without feedback}
Without feedback, the system follows the ubiquitous cooling catastrophe
picture. Baryons condense in the halos and rapidly radiatively cool
 due to their high density. A disk galaxy is formed in the center of the
high resolution region, with a (baryonic) mass of $1.2\times10^{10}$
\msol. The disk exhibits a (visibly striking) cutoff in particle density
at a radius of 8 kpc, and similarly, the vertical distribution of the disk
falls off abruptly above and below 1.5 kpc of the equator. A double
exponential fit of the gas density profile (see section \ref{tdprofiles})
clearly displays the rapid fall-off in density with radius beyond 8 kpc.
Star formation proceeds rapidly due to the high density, and is initially
concentrated in the nucleus of the disk (which contains 40\% of the
baryonic mass and has a 0.6 kpc diameter\-much smaller than the softening
length). Stars formed in the nucleus diffuse away from it, forming a
stellar bulge approximately 3.0 kpc in diameter (compare the radial
density profiles in \fig~15). Due to the low resolution, the
hierarchical formation of the disk is represented poorly, with only a
handful of progenitors merging to form the disk.

At late times $z=1.09$, the disk exhibits a number of features that have
been observed previously. There is a deficit in angular momentum, with the
specific angular momentum of the baryons corresponding to that of an
elliptical system for the given mass scale. Consequently the disk has a
small radial extent. At $z=1.01$ the disk undergoes a major merger with
another system of mass $6\times10^{10}$ \msol\ (in baryons)  at a speed of
300 \kms\, relative to the center of mass frame for the major disk. As is
generally observed in simulations with stellar and gaseous components, the
resulting morphologies of the gas and stars differ significantly. The gas
cores merge, creating a very dense core while the stellar components
merge, producing `shells' as observed in ellipitical galaxies (Quinn
1984). A tidal tail is also produced during the merger and is populated by
both gas and stars.

\subsection{Star Formation Rate}
Unfortunately, even though the SFR normalization was adjusted to 0.025,
the SFR in the simulations appears to be somewhat low. Although the plots
in \fig~\ref{cosmo.SFR} show SFRs in excess of 70 \msol yr${}^{-1}$, it
should be noted that this value is integrated over $8\times10^{11}$ \msol.
Diagnostics from the simulation indicate that of this mass,
$6.5-7\times10^{11}$ \msol\ is not involved in star formation ($T > 30000$
K or $\rho<\rho_{sf}$). Beyond the main disk and the merger companion (a
combined baryonic mass of $1.8\times 10^{11}$ \msol\ of which 60\% is in
star forming regions), tertiary halos contribute only $2.4\times10^{10}$
\msol\ of star forming matter. Hence, the bulk of the SFR is derived from
the main disk and its major companion. It should be emphasized that the
halo correspondence between simulations is not perfect, but given the
difficulty in accurately calculating the cooling rates at low resolution,
and the highly non-linear nature of the dynamics, this is not surprising.
Well-defined halos, \ie those formed with 500 or more particles, do
correspond well, as can be seen in the radial plot in
\fig~\ref{cosmo.lx2r}. There are also small synchronization errors ($10^5$
years) between the analysed time-step outputs.  To examine the effect of
changing parameters a number of auxiliary simulations were run, the
details of which are discussed in section~\ref{auxsim}.

\subsection{Effect of feedback on SFR and morphology}
The most noticeable difference in the ensemble is that the temperature
smoothing version does not lead to a significantly different final
structure. This is contradictory to the isolated results where temperature
smoothing is seen to promote violent winds and disk disruption. The reason
for this is that the density of the first objects is so high, \\

{\epsscale{0.95}
\plotone{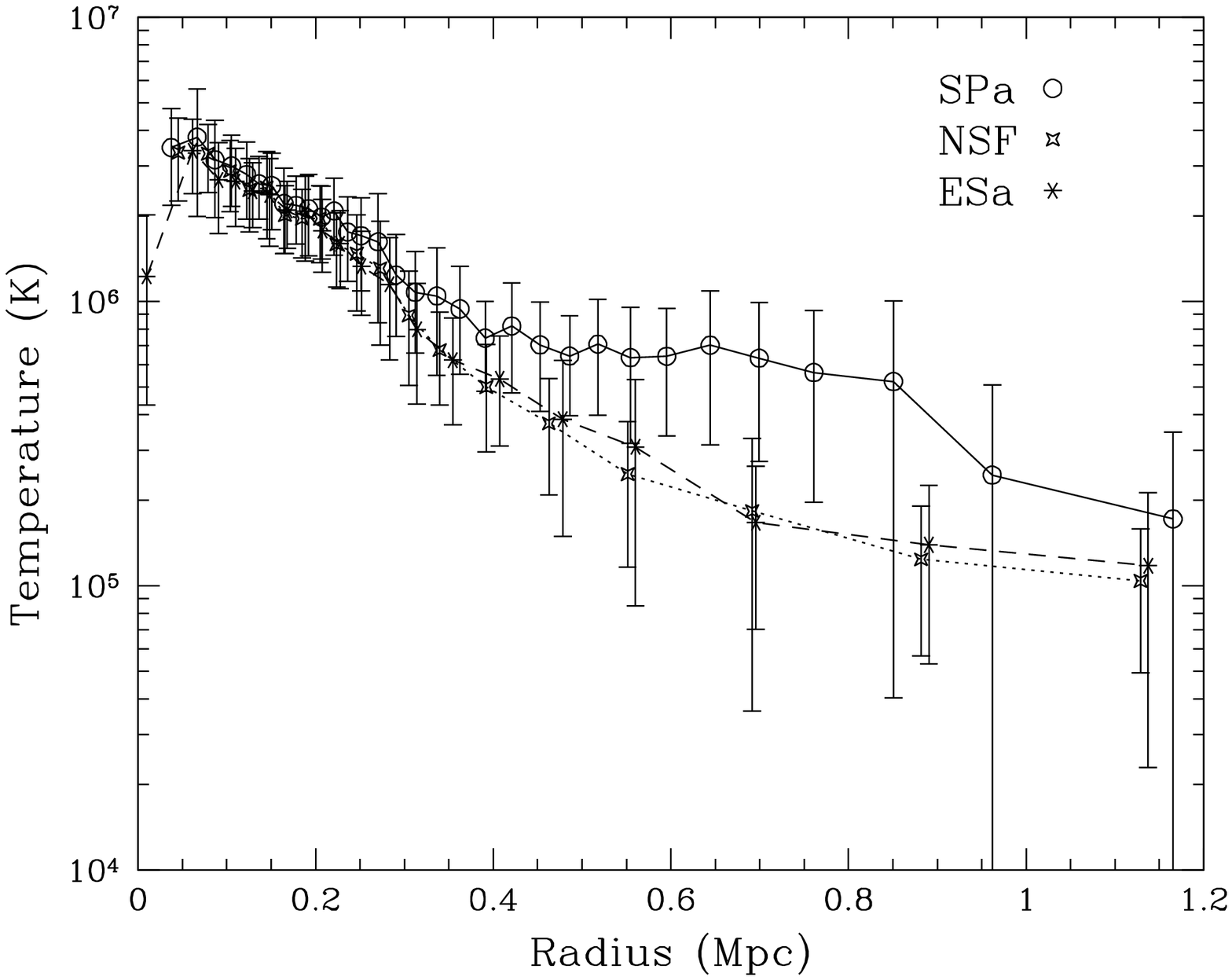}}

{\small {\footnotesize {\sc Fig.}~14.}---Radial temperature profile for
the NF, SPa and ESa runs.  The error bars denote 1$\sigma$ variation about
the bin mean, with the data plotted in Lagrangian bins of size 208
particles. The SPa and SPna (not shown)  profiles both exhibit a higher
temperature at large radii (see text). Of the remaining algorithms, all
follow profiles similar to the ESa and NF simulations.} \bigskip

\noindent $n_H >
1\;{\rm{cm}}^{-3}$, that the cooling time ($\simeq 0.1$ Myr) is short
enough to remove the SN energy within a time-step, unlike the
isolated simulation where the resolution is high enough to allow a
reduction in density due to expansion of the feedback region and
consequent reduction in the cooling time. To test what happens when the
feedback energy is allowed to persist, the TS simulations were run with
the adjusted cooling mechanisms. As expected, these simulations produced
more diffuse structures. 

At z=1.09, the morphology of the major disk was examined. Without
exception, all the simulations produced a disk with a clearly defined
cutoff radius of $9.2^{+2.1}_{-1.6}$ kpc ($>2h_{min}$). This result is the
same as the NF run. However, models including feedback were fatter at the
disk edge, (the TSa run with a thickness of 5 kpc, being 30\% wider than
the NSF run). `Bubbles' were noticeable in the disks, more so in the ESa
run than others because the TSa and TSna runs were already quite diffuse. 
Feedback did not change the radial extent of the disk which suggests that
it must be determined largely by the dark matter potential.  It cannot be
due to the central concentration of baryons, since the TSa and TSna runs
effectively destroy this concentration yet still have approximately the
same radius. 

While the internal structure of the major disk was not significantly
different across all simulations, that of the merged system was. For the
NF simulation, the gaseous cores were much more tightly bound than those
in the TSa and TSna runs, but were largely similar to those in the ESa,
Esna, SPa and SPna runs. In particular, because the gas cores are
sufficiently inflated in the TSa and TSna runs, the gas undergoes a smooth
merger, and for the TSa run the feedback is sufficient to produce a {\em
disk} as the result of the merger. Note that the stellar components evolve
in a similar fashion though, producing shells, and a widely dispersed
final stellar structure. 

The SFRs for the main simulations are plotted in \fig~\ref{cosmo.SFR}. 
The upper left panel shows the results for the simulation without
feedback, and gives an illustration of the smoothing effect of the 160
step running average used to smooth the data. All algorithms agree on the
early SFR, which reaches 1 \msolyr\ at $z=3.9$, since sufficient time must
pass for the star mass of particles in the highest density regions to
reach the mass threshold for creating a star particle (the first star
particles are created at $z\simeq3$). At late times, the merger causes a
strong star burst which is visible in all of the SFRs, albeit somewhat
suppressed in the simulations with strong feedback. The relative effect of
the different feedback algorithms was compared by calculating the
reduction in the cumulative mass of stars at $z=1.09$, as a percentage
relative to the no feedback run (total $9.7\times10^{10}$ \msol).  The
algorithms with the most significant effects are, in order, TSa
(25\%),TSna (23\%), SPa (18\%)  SPna (13\%) and TS (10\%) while the energy
smoothing variants ESna (3\%), ESa (2\%) have comparatively little effect
on the SFR. At earlier, epochs, in particular shortly after $z=2$, both
the TS and SP runs have a significantly higher reduction in the SFR. For
example at $z=1.6$ the SPna run has an SFR only 20\% of the NF run. As in
the isolated simulations, the SP algorithms eject particles due to
asymmetry in the local distribution of particles and the subsequent
reduction in the gas density is the main source of the SFR reduction over
the energy smoothing variants. 

\subsection{Halo profiles}\label{tdprofiles} 
In view of the results from the isolated simulations, the
halo structure was examined to see if there was any difference
between algorithms. 
\Fig~14 compares the gas halo temperature for the NF, ESa
and SPa runs. The higher temperature seen at the edge of
the SP profile (beyond 200 pc), is difficult to attribute just to `hot'
particles being ejected to that radii. Tracing the orbits of a
number of ejected particles showed that the largest distance they
achieve from the core is 150 kpc. It is noticeable that at a radius of 150
kpc, the temperature of the SP feedback halos is higher than
that of the others. A plot of the radial pressure showed that beyond
200 kpc, the pressure in the SP halos is a factor of two
higher than in the other runs. A plot of the cumulative density
versus
radius confirms that more of the gas lies
at large radii (beyond 200 kpc) for the SP feedback. This
indicates that the particles ejected from the disc by SP feedback are
acting like a piston on the outer regions of the gas halo, subsequently
leading to higher
temperatures in the infalling matter at large radii (since the
gravitational compression remains dominated by the dark matter). 

The density profiles for the baryons can be fit by a double exponential,
with the break between the two profiles occurring at the edge of the disk. 
An argument can be made that the presence of the gas/stellar core suggests
that the disk should also exhibit a double profile; however the structure
is sub-resolution. In particular, when the smoothed density is examined
(which is used in the SFR calculation), there is very little difference
between simulations. A summary of least squares fits for the inner and
outer parts of the density profiles is given in table~\ref{tbl-3}. The
fitting was somewhat arbitrary since the break between the fits is decided
by eye. Note that for the TS \\

{\epsscale{0.95}
\plotone{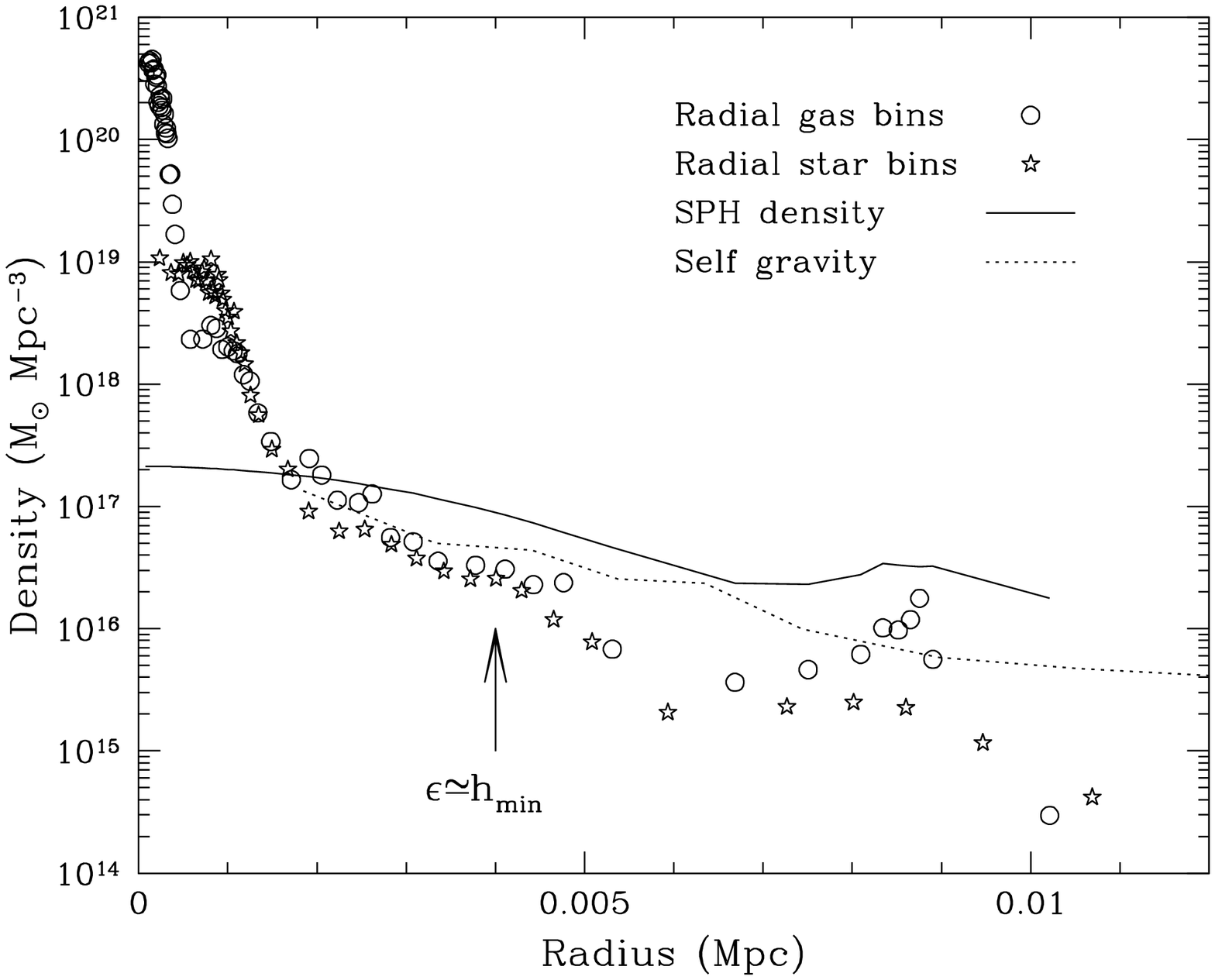}}

{\small {\footnotesize {\sc Fig.}~15.}---Density profile for the NF run. 
Spherical Lagrangian bins (52 particles) have been used to bin the star
and gas data.  Clearly the gas and stellar data are comparatively similar.
The stellar bulge has roughly constant density and extends to a radius of
1.5 kpc, the gas nucleus extends only just beyond 0.5 kpc. The SPH density
data is a radial binning of the raw SPH density values which, because of
smoothing, do not increase to the exceptionally high values seen in the
radially binned stars and gas. Further, it reflects the 2-dimensional
density better than the spherical bins. The self-gravity line corresponds
to 0.4 times the spherically binned dark matter density.} \bigskip

\noindent variants, this was particularly difficult since the transition
from disk to halo is less clear, \ie the density curve is smoothly
decreasing as opposed to a sharp discontinuity visible in the other data
sets. The inner fits, which give an effective scale length for
3-dimensional baryon distribution in and about the disk, are broadly
similar and are given by $s_L=0.75^{+0.25}_{-0.16}$ kpc (ignoring the
auxiliary simulations presented in section\ref{auxsim}).  Note that the
gas cores tend to distort the fits toward shorter scale lengths and the
three-dimensional profile underestimates the scale length that would be
interpreted from a surface density plot. There is a clear trend for the
adiabatic feedback schemes to have longer scale lengths than the
non-adiabatic versions. This is to be expected since the adiabatic
feedback keeps the gas more diffuse. The outer fits are more problematical
since satellites severly distort the radially averaged density profile. 
Given the sensitivity of the slope to these perturbations, it is difficult
to draw detailed conclusions from these data.

The gas and stellar density profiles are broadly similar since the stellar
disk evolves out of the gas. Star particles that are formed within the
dense central gas core eventually orbit at a larger radii than the parent
particle since they are not affected by the viscous forces felt by the
gas.  A comparison of the gas to stellar density profile is shown in
\fig~15. The smoothed gas density (\ie the SPH density) is
remarkably similar across all simulations indicating that the SFR should
be similar (modulo the effect of feedback events). The clear rise in the
density at small radii is a signal of the gas core, albeit at
sub-resolution scales. For the TSa/na runs this core was removed due to \\

{\epsscale{0.95}
\plotone{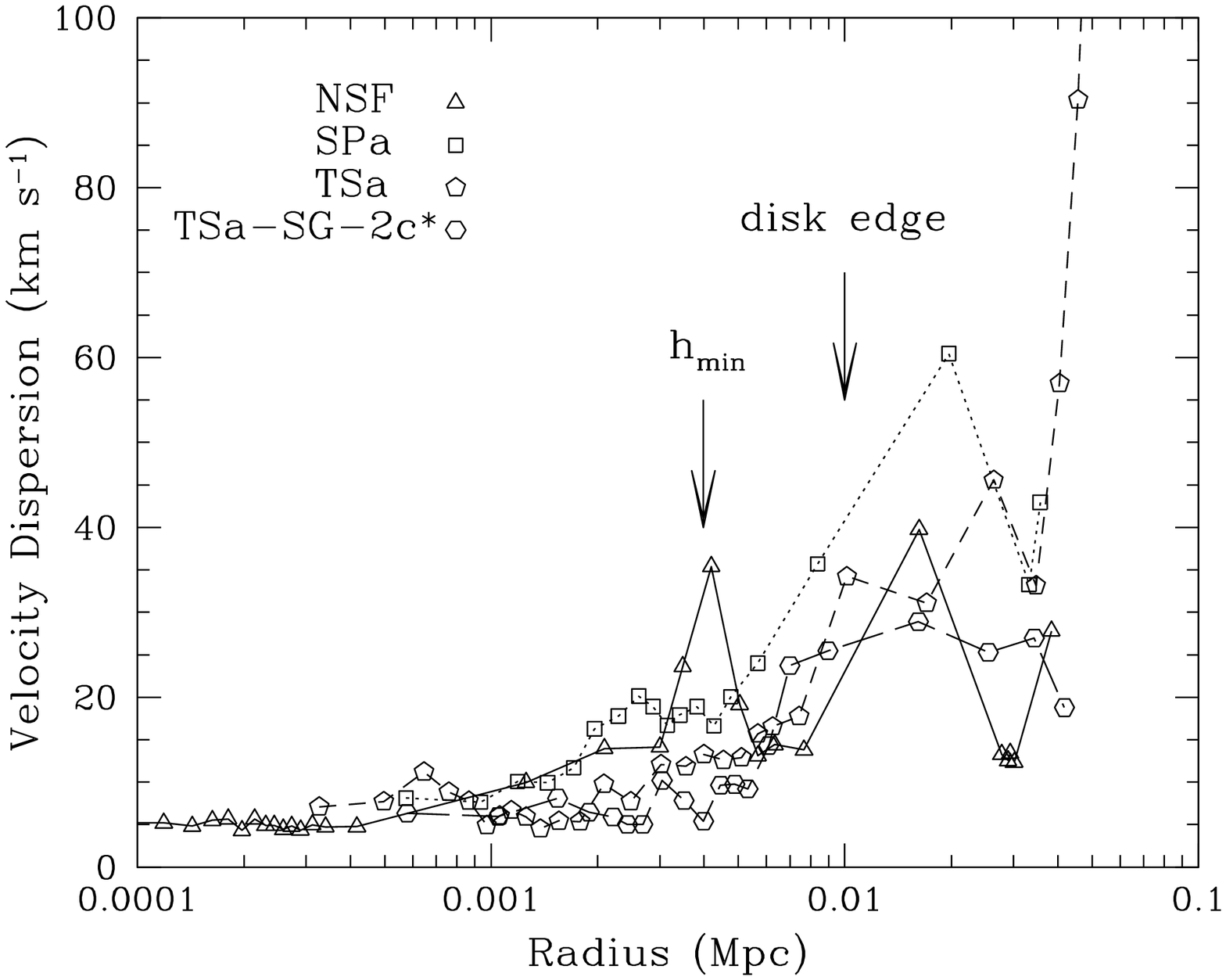}}

{\small {\footnotesize {\sc Fig.}~16.}---Radial velocity dispersions in
the gas disk. The minimum smoothing length $h_{min}$ and disk edges are
marked for clarity.  In contrast to the isolated simulations, there
appears to be no correlation between more violent feedback producing
higher velocity dispersion. This is partially due to the fact that merging
dwarfs produce a far higher velocity dispersion than feedback, and also
the disks are not well resolved.} \bigskip

\noindent the strong feedback. For the SP runs, the ejection of particles
also
lowered the core mass. The ES runs were incapable of inflating the core
once formed. 

The density profiles for the dark matter differ little from simulation to
simulation ($r_{200}$ differs across all simulations by only 1\%). At
least with this mass resolution, there is no evidence for feedback being
capable of rearranging the dark matter structure (\cite{NEF96}). 

\setcounter{figure}{16}
\begin{figure*}[t]
\epsfxsize=14cm
\centerline{\epsffile{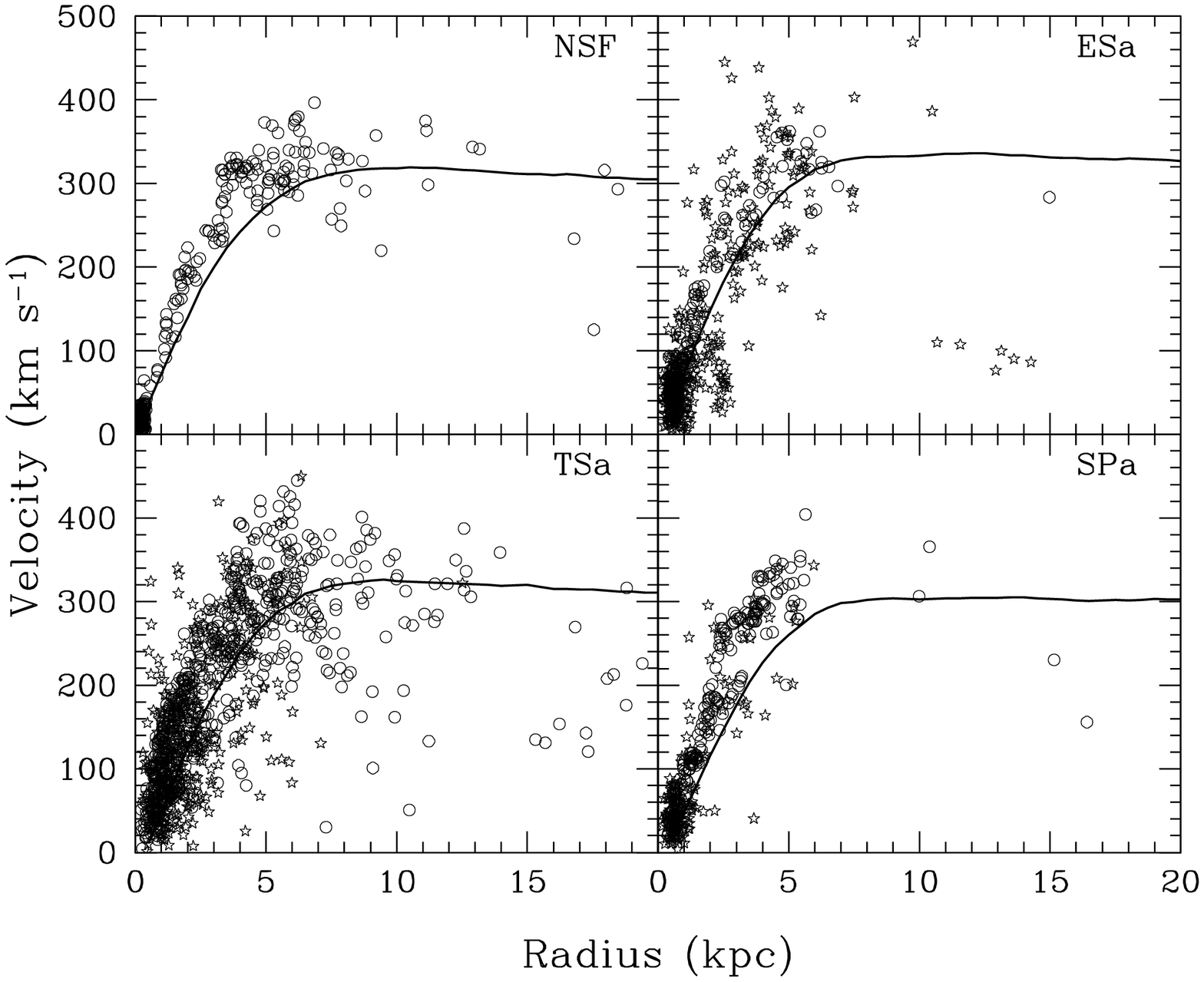}}
\caption[Rotation curves for the NSF, TSa, ESa and SPa runs in
the disk formed in the cosmological simulation.]{Rotation
cruves for the NSF, TSa, ESa and SPa runs. The solid line shows
the Plummer softened rotation curve while points indicate the tangential
velocity of individual particles. Circles represent gas particles and open
stars stellar particles. Particles plotted lie in a band 10 degrees wide
about the plane perpendicular to the angular momentum vector of the gas.
More particles appear in the TSa plot since the dense stellar core is
inflated while in the other plots most of these particles fall
outside the selected plane (since they are contained in the dense core).
The integrated rotation curves are broadly
similar and the particle data differ only marginally, with more
velocity dispersion being visible in the TSa run.} 
\label{cosmo.rc}
\end{figure*}

\subsection{Rotation curves and angular momentum}
\Fig~\ref{cosmo.rc} displays the (Plummer softened) rotation curves,
\begin{equation}
v^2_{c}(r)={GMr^2 \over (r^2+\epsilon^2)^{3/2} },
\end{equation}
and particle tangential velocities for the run without star formation
compared to the ESa, SPa and TSa runs.  For all simulations, the
tangential velocities rise more sharply than the rotation curve. However,
the initial slope of the rotation curve is dominated by the softening
parameter and a 12\% reduction in the softening length is enough to fit
the tangential data, hence this should not be considered a significant
discrepancy. The 300 \kms\, peak of the rotation curve is consistent with
the mass of the disk and halo, although the SPa run is slightly lower
since gas has been ejected out of the disk into the halo. 

The TSa run shows significantly increased dispersion in the tangential
velocities because of the feedback, but does not have a larger disk
diameter (in keeping with the similar scale lengths found in the analysis
of the density profiles). Since particles involved in feedback regions
tend to be ejected vertically from the disk, \ie preferentially into low
density regions, a large tangential velocity dispersion can arise. This is
seen most clearly in the TSa plot.

In view of the rotation curves being similar across all simulations, it
would be expected that the velocity dispersions should also exhibit
similar profiles. A comparison plot of the NSF run compared to the SPa,
TSa and TSa-SG-2c* run (note this run has double the fiducial SFR and no
self-gravity criterion, see section~\ref{auxsim}) is shown in
\fig~16. As expected, the profiles are broadly similar with a
maximum difference between the plotted runs of 20 \kms\ at the disk edge.
It is interesting to note that the TSa run has a large tangential velocity
dispersion, as is visible in \fig~\ref{cosmo.rc}, but a comparatively low
dispersion in the radial direction. This is a result of feedback
preferentially ejecting particles vertically, in turn boosting the
tangential velocity component more than the radial. The data for the NF
run (not shown) are dominated by an ongoing merger which introduces a very
large dispersion (120 \kms) at the edge of the disk, far larger than that
produced by any feedback. The NF and ESa runs also show the effect of the
merger, with peaks in the data at 4.2 kpc and 3.6 kpc respectively,
corresponding to the position of the strongest perturbation within the
disk. The SPa and TSa runs are less affected by the merger since the dwarf
has been reduced in mass by the stronger feedback. It is clear that the
low resolution in the disks and the complications of ongoing mergers make
it difficult to draw conclusions from this data. 

To see how much angular momentum has been lost by the disk, the specific
angular momentum of the cores is compared to that of the dark matter halo
(within $r_{200}$) in \fig~18. The angular momentum for the
halo gas (the gas for which $\delta <2000$ within $r_{200}$) and that of
the stellar component of the disk are also shown.  For all simulations,
the disk system shows a deficit of specific angular momentum when compared
with the dark matter. By breaking the disk into its stellar component and
gas component, it becomes clear that in the simulations with feedback that
are shown (TSa, SPa and TSa-SG-2c${}^*$) there is a {\em trend toward
higher angular momentum values for the gas disk}. The highest value, that
from the TSa-SG-2c${}^*$ simulation, just falls within the disk region of
the parameter space. However, the stellar disks all fall in the elliptical
region of the parameter space, and the effect of feedback on them is
small. Note that the purely gaseous run also sits in the elliptical region
and the gas disk in the no feedback run has marginally higher specific
angular momentum than the stellar component. The NF run is misleading,
since a merger is going on at the edge of the disk leading to higher
angular momentum values as compared to the other feedback runs (although
by the end of the merger the opposite \\

{\epsscale{0.95}
\plotone{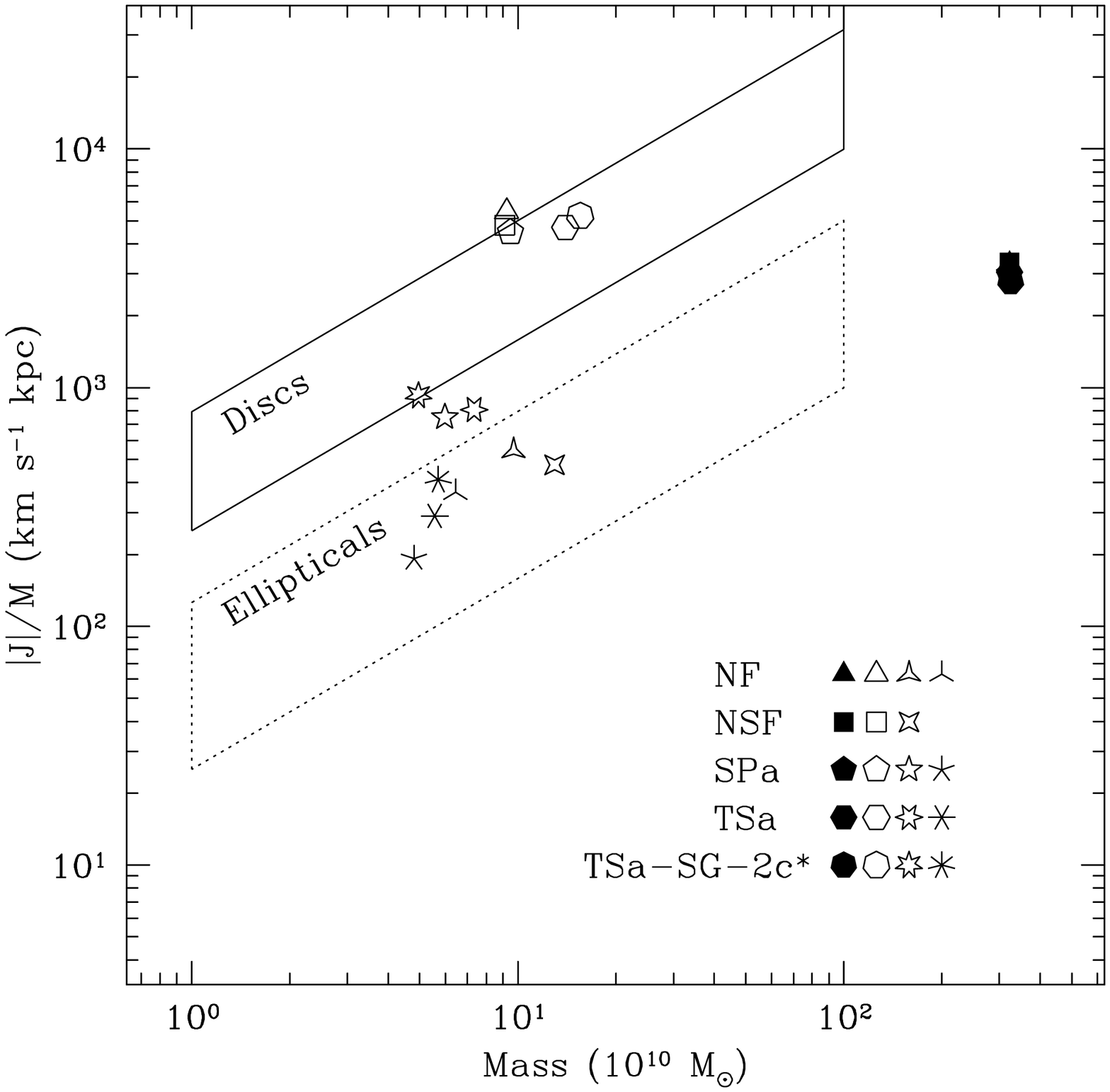}}

{\small {\footnotesize {\sc Fig.}~18.}---Specific angular momenta versus
mass for different components of the system for a number of different
feedback algorithms. The filled polygons plot the angular momentum of the
dark matter within $r_{200}$, the open polygons that of halo gas (all gas
that does not fall above $\delta=2000$), pointed stars that of the gas in
the main disk ($\delta>2000$) and finally the centrally connected stars
show that angular momentum of the stellar component of the disk. The runs
with feedback show a small but noticeable trend toward higher angular
momentum values for the gas disk component (contrast with the NF run).
Both dark matter and gas halo values are in broad agreement as expected.}
\bigskip

\noindent will be true due to core-halo interaction).  For all
simulations, the angular momentum of the halo gas is larger than that of
the dark matter since the dark matter plot includes the contribution of
the dark matter core, which has little angular momentum but a significant
amount of mass. It is clear from the plot that if the halo gas were to
fall smoothly onto the disk, then it should be possible to form a disk
with an angular momentum value midway between that of the halo gas and
disk system. Note infalling satellites may still disrupt the disk and
cause still further angular momentum loss. 

\setcounter{figure}{18}
\begin{figure*}[t]
\epsfxsize=15cm
\centerline{\epsffile{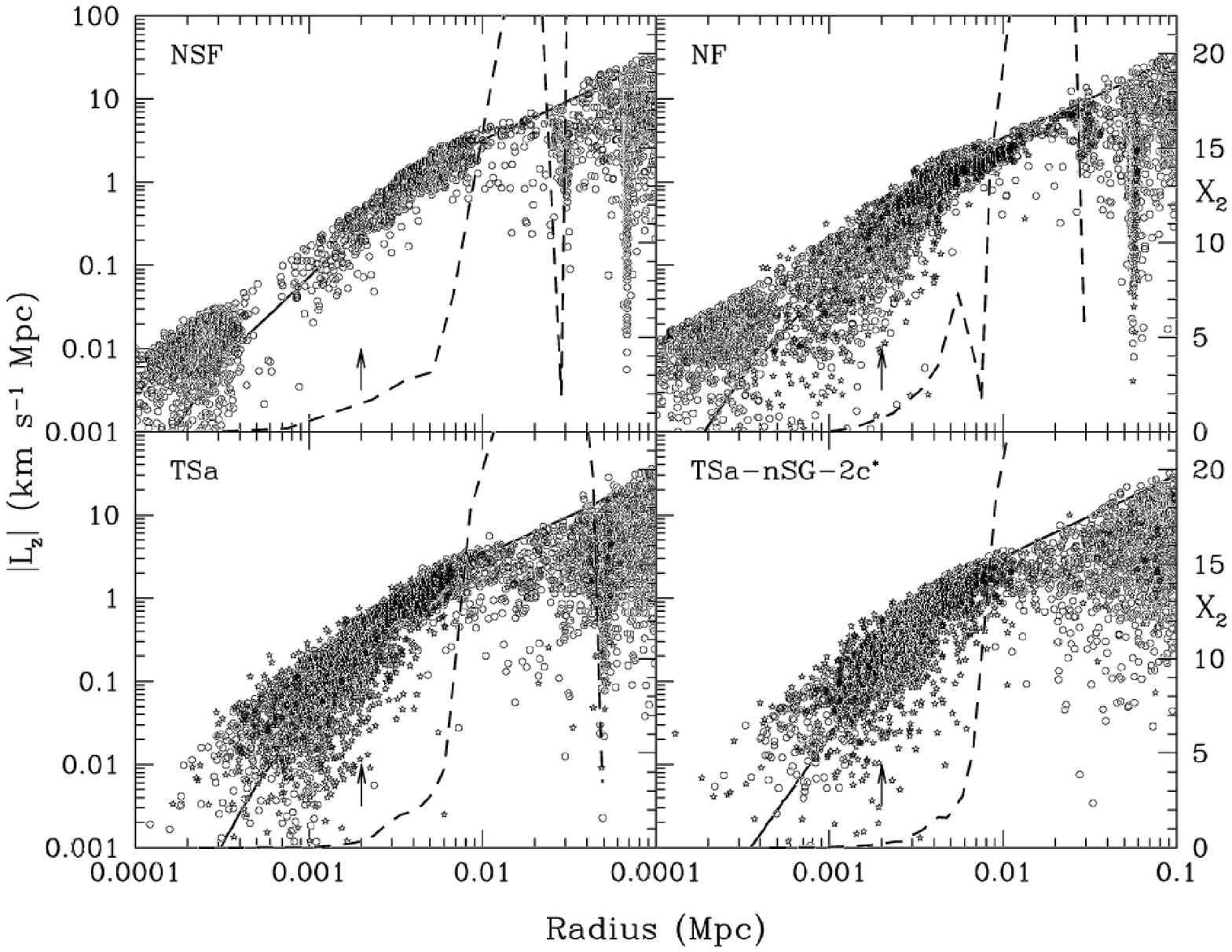}}
\caption[Plot of specific angular momentum versus radius, showing the raw
particle data versus $|L_z|$ calculated from the rotation curve at
$z=1.09$]{Plot of specific angular momentum versus radius, showing the raw
particle data versus $|L_z|$ calculated from the rotation curve at
$z=1.09$ (counter rotating particles are not shown). The
$X_2(R)$ stability parameters for the disks are also shown. The raw data
fits the expected
curve well, indicating that at least at the current epoch (prior to
significant accretion) there has not been significant angular momentum
loss as a result of disk formation (a bar is yet to form). The large
concentration of matter at very small radii in the NF and NSF runs is the
central core and is slightly offset, since the center of mass velocity was
measured over the entire disk rather than the core. The $X_2(R)$ plots are
all similar with little difference even for the extreme feedback in the
TSa-SG-2c${}^*$ run. Disk stability occurs at $X_2(R)\simeq3$ and is
achieved
at the smallest radius of 3.5 kpc in the NSF run and at the largest radius
of 6.0 kpc for the TSa-SG-2c${}^*$. Note that the TSa runs provide
sufficient feedback to remove the central mass concentration, but fail to
increase the disk radius significantly.}
\label{cosmo.lx2r}
\end{figure*}

The z-component of the specific angular
momentum {\bf L} is shown in \fig~\ref{cosmo.lx2r}. If significant angular
momentum loss occurs as
a result
of disk formation then a larger proportion of the disk mass should lie
beneath the line formed by $|\bf{r}\times \bf{v}_{circ}|$. Since this is
not the case, it is clear
that at $z=1.09$ there has been little angular momentum loss (within the
disk) due to bar formation. However, all the disks are deficient in
angular momentum relative to the halo, since they have been formed in a
hierarchical process which is subject to core-halo angular momentum
transport. Remarkably, the $X_2(R)$ stability parameter (\cite{T81,BT87})
plots
are all
similar, with all the disks achieving the $X_2(R)\simeq3$ stability
requirement
just beyond the 4 kpc softening radius.  As was shown by
Dominguez-Teneiro \etal (1998), it is more than
likely that if the baryonic mass is redistributed into an exponential
disk,
then the $X_2(R)$ parameter for this system does not achieve stability
until a much larger radius, \ie the cores provide disk stability. Note
that the kink in the $X_2(R)$ plot for the NF run is due to the merger
previously discussed and it does not affect the disk stability greatly
(the kink drops to 2 at 6.5 kpc but quickly returns to stable values). 

\subsection{Auxiliary simulations}\label{auxsim}
To adequately examine the parameter space of these simulations and
also determine the effect of SPH algorithm changes would take
an excessively long time. Alternatively, by conducting a few auxiliary
simulations, much can be learnt about the outer edges of
the parameter space. To understand what happens when the SPH algorithm is
changed, in particular in relation to the treatment of high density
regions, it is simple to contrast to one of the previous simulations run
with the same parameters.

\begin{figure*}[t]
\epsfxsize=16cm
\centerline{\epsffile{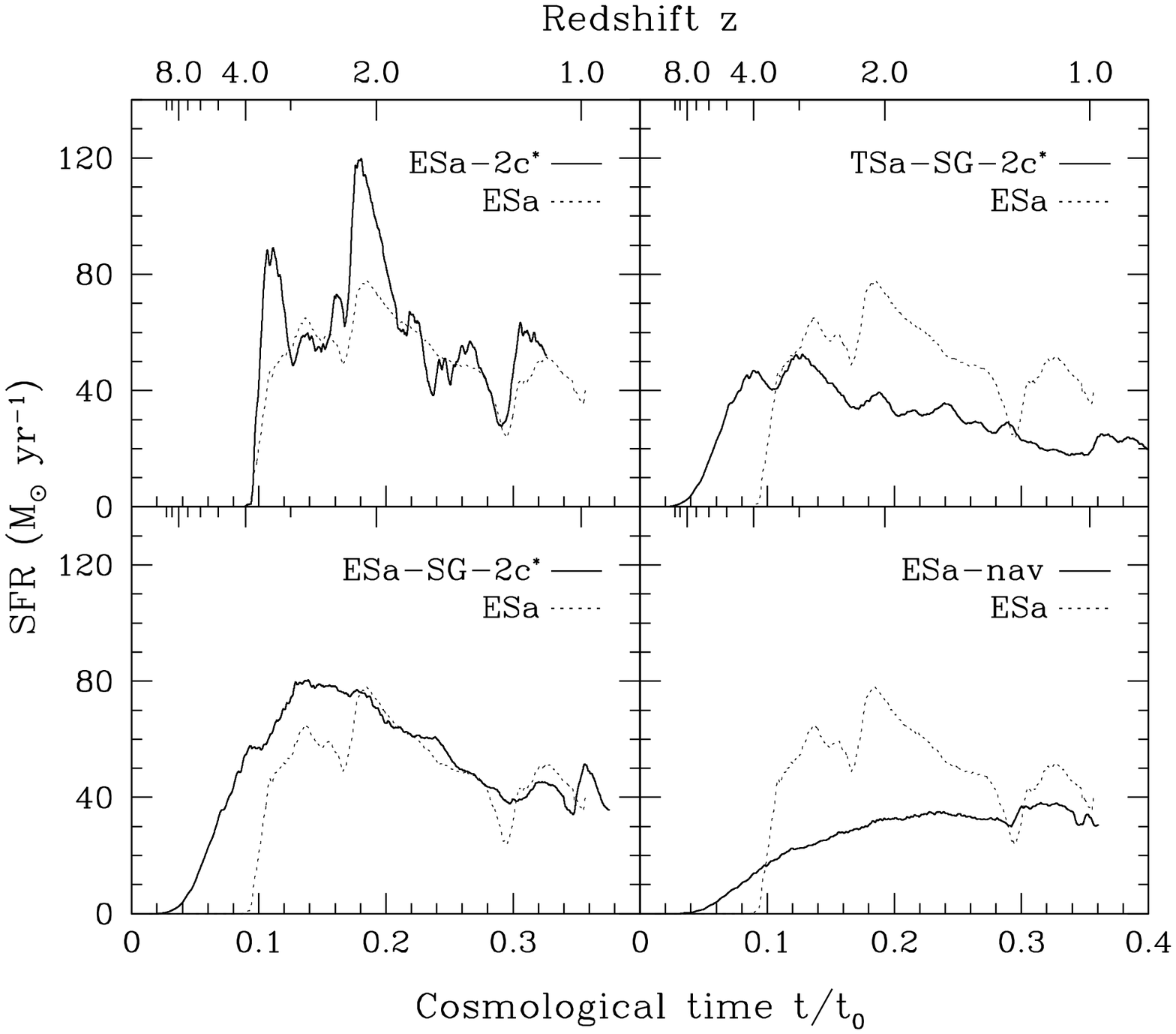}}
\caption[SFRs for the auxiliary cosmological simulations.]{
SFRs for the auxiliary cosmological simulations.  The SFR shown is
integrated over the entire gas sector of the simulation ($8\times 10^{11}$
\msol), and time averaging is used to smooth the data. Doubling the SFR
normalization does not lead to a perfect doubling of the SFR, although
peak
values are close. Removal of the self-gravity criterion leads to the
very early onset of star formation. The TSa-SG-2c*
simulation has
strong feedback and even with the higher SFR normalization still has a
lower peak SFR than the ESa run (plotted for comparison). The ESa-SG-2c*
run, while having a peak SFR earlier, is not significantly different from
the ESa run. Changing the short range behaviour of the SPH solver
leads to a markedly different SFR, as shown in the plot of the ESa-nav
data. The peak of the SFR is later and lower, and the self-gravity
criterion only mildly delays the onset of star formation. } 
\label{aux.SFR}
\end{figure*}

As indicated in previous sections, to determine the effect of removing the
self-gravity criterion, the ESa simulation was rerun without it. This
simulation is denoted ESa-SG. Also the effect of doubling the star
formation rate normalization was tested in a simulation denoted ESa-2c*.
Since most of the simulations conducted appeared to be relatively
unchanged by the introduction of feedback, a simulation with extremely
violent star formation and feedback was run (twice the fudicial SFR
normalization with TSa feedback and also without the self-gravity
criterion).  This simulation, denoted TSa-SG-2c${}^*$, is not particularly
realistic (star formation begins very early and the feedback is
over-efficient)  but it does provide an excellent guide to the limits of what
feedback can accomplish.  Because of the strong feedback, the formation
of dense gas cores was all but prevented, with the exception of very large
$>5\times10^{10}$ \msol\ systems. Hence it was possible to integrate the
system to later times since the SPH algorithm did not suffer significant
slow down. An ESa-SG-2c${}^*$ simulation was also run to contrast with the
temperature smoothing version. 

As an alternative to smoothing over all the particles within the minimum
smoothing length, $h_{min}$, literature on galaxy formation (this appears to
be implied
in \cite{NW93,BB97}) suggests that one may continue
search over $N_{smooth}$ neighbour particles. Such a
procedure places a direct limit on the maximum resolved density: the
volume normalization is set by $h_{min}$ and the summation over neigbours
is limited to $N_{smooth}$ particles. 
In turn, this sets a bound on the
maximum SFR per particle and consequently within the system as a whole. It
also sets a bound on the cooling rate. The reasons for making this change
are primarily related to efficiency: if a simulation smooths over all the
particles within $h_{min}$ it can exhibit a severe slowdown. For
example, nearly all the simulations without this adjustment slow down by a
factor of 7 from start to finish. Hence to examine the effect of making
this change, the ESa simulation was re-run with this high density
treatment.  This simulation is denoted ESa-nav. It is interesting to note
that this method can be made to almost exactly reproduce the density
values calculated in a simulation with $h_{min}\rightarrow 0$, a volume
normalization is the only thing that need be applied.

In the ESa-SG run, the effect of removing the self-gravity criterion is
that star formation begins at a very early epoch (1 \msolyr\  at
$z=8.3$, see \fig~\ref{aux.SFR}).  It is initially confined to a few
particles
(recall the
$\nabla.{\bf v}$ criterion must also be fulfilled) leading to an SFR of
0.2 \msolyr.  During later evolution, the SFR is only marginally lower
than that of the ESa run, the largest difference being 10 \msolyr\  at
$z=2.1$ (a 12\% difference). Further, the disk and halo structure are
comparatively similar, as measured by density and temperature profiles,
and the dense gas core is still formed. 

The plot of the SFR in the ESa-2c${}^*$ simulation (\fig~\ref{aux.SFR})
shows that  doubling the SFR normalization leads
to a stronger
initial star burst as the gas overcomes the self-gravity criterion, but
does not produce an SFR that is exactly double that of the standard
simulations. The peak SFR of 120 \msolyr\ at $z=2.1$ is 56\% higher than
that in the ESa run (77 \msolyr). The SFR is not simply doubled because of
both increased feedback and the finite amount of gas available for
star formation. Notably the dense gas core was still formed, showing that
even with the increased SFR, ESa is still incapable of producing a
significant effect on morphology.

As expected, the TSa-SG-2c${}^*$ simulation leads to markedly different
results than any of the previous simulations. Because of the extreme
feedback and consequent absence of small progenitors, the disk assembly
process is very smooth. There is essentially no formation of a gas and
stellar nucleus within the disk. Star formation begins at the same epoch
as the ESa-SG run, albeit at a higher rate due to the increased SFR
normalization which yields 1 \msolyr\ at $z=9.1$. At $z=6.2$ the SFR
reached 15 \msolyr\ and a peak of 52 \msolyr\ occurred at $z=3.0$. The
most
noticeable difference in the SFR is that at late times it falls off
precipitously. At $z=1.0$ the SFR is 18 \msolyr\ compared to 
36 \msolyr\ for the ESa run and by $z=0.5$ it had fallen to 5 \msolyr.
Note the decay
in the SFR versus time was linear rather than exponential, This has been
observed before in parameter space explorations (\cite{casca}). The less
energetic feedback provided by energy smoothing leads to the
ESa-SG-2c${}^*$ simulation producing a much higher peak SFR, namely 80
\msolyr\ at $z=2.7$. In keeping with all the other energy smoothing
simulations, a central gas nucleus was formed in the disk and the
temperature and density profiles remain similar to the ESa run.

At $z=0.5$, the TSa-SG-2c${}^*$ disk was analysed to see if the accretion
of the halo gas had indeed allowed the formation of a disk without an
angular momentum deficit. By this epoch, the disk had grown to a diameter
of 12.0 kpc, which is 17\% larger than the value from $z=1.09$ and the
mass had increased by 30\% to $1.39\times10^{11}$ \msol. $r_{200}$ had
grown to 285 kpc. The ratio of the specific angular momenta for the gas
core and dark matter, increased to 0.40 (20\% increase). Visualization
of the system shows that the ratio cannot increase significantly as there
are very few cold gas clumps within the dark matter halo available for
accretion. Most have been blown apart by feedback. Note that at the center
of the halo the gas density is approximately $n_H\simeq10^{-3}$ cm${}^-3$
and the temperature is over $10^6$ K, leading to a cooling time of greater
than $4\times10^9$ years. Hence very little of this gas, which has a large
specific angular momentum, can cool on to the gas disk before $z=0$. 

The results from the ESa-nav simulation are very different. The peak SFR
is 38 \msolyr\ and it occurs at a much later time than the rest of the
simulations ($z=1.1$). Also star formation begins slightly later than that
seen in the simulations without the self-gravity criterion (1 \msolyr\ by
$z=6.9$) which is significantly earlier than the standard simulations (1
\msolyr\ by $z=3.9$). The reduced SFR is due to the density values
calculated by this method being lower and not due to any increased effect
of feedback. The change in the epoch at which the self-gravity criterion
is overcome is due to the change in the search radius. In the center of
the halo the dark matter profiles have a much shallower density profile
than the baryon cores. When the neighbour search is conducted over the
reduced radius only the core is sampled  rather than the full 4 kpc
radius, \ie the sharp decline in baryon density at the disk edge is
ignored. The disk formed is broadly similar to that in the ESa run,
although it has a slightly larger baryon core, a thinner structure (2 kpc
thick) and a smaller radius. 

As was expected, the wall-clock per
time-step slowdown was less severe for this code and it was roughly double
the speed of the standard implementation.  Note that this is an
underestimate of the efficiency since the $h$-update algorithm did not
accurately calculate the required change in the search radius for regions
where it was less than $2h_{min}$, see \cite{rob} for a full discussion.
The algorithm preferentially smooths
over too many particles, and an accurate calculation would require a very
short time-step. 

\section{Summary and Discussion}\label{conc2}
This paper details a study of a number of different feedback
algorithms, comparing the effect on high resolution isolated systems and
low resolution hierarchical simulations. The parameter
space of the model was explored as were the effects of small changes in
the hydrodynamic solver.  

Principal conclusions follow:
\begin{enumerate}

\item As would be expected on energy budget grounds, the temperature
smoothing (TS) feedback algorithm has the most impact on structure
formation.
Single particle (SP) feedback has a less significant but still noticeable
effect and it produces a distinct change in the temperature profile of the
halo at large radii. Energy smoothing (ES) is the least effective of the
three fundamental mechanisms. 

\item Feedback can be shown to have a large effect in systems that are
well resolved in terms of particle number. In particular, the Milky Way
and
NGC 6503 prototypes are strongly affected by feedback even though the
softening lengths were chosen to be comparable with those of cosmological
simulations. Both `blow-out' and `blow-away' can be produced, depending
upon the feedback algorithm. 

\item The NGC 6503 prototype is more strongly affected by feedback than
the Milky Way prototype, \ie feedback has more effect on low mass systems,
as expected. In the NGC 6503 model it is possible to
differentiate between the cooling mechanisms.

\item In hierarchical simulations even an excessive amount of feedback,
TSa-SG-2c${}^*$ for example, produces little effect on the properties of
large ($>10^{11}$ \msol) disk galaxies at early epochs. Rotation curves
and density profiles remain broadly similar. 

\item Although small at the mass scale probed by the hierarchical
simulations, there is a distinct trend toward higher specific angular
momentum values in the disk with increased feedback. At lower mass scales,
equivalently at higher resolution, the effect should be more noticeable.

\item The revised cooling mechanism has little effect for the energy
smoothing algorithm since insufficient energy is input, and consequently,
the estimated density does not lower the cooling rate significantly.
Conversely, for single particle feedback and temperature smoothing, the
energy input is more than sufficient to force the cooling rate from the
estimated density to be much
longer than the local one. 
 
\item Morphologies remain broadly similar in hierachical
simulations although feedback can reduce the SFR, compared to runs without
it, by over 25\% at z=1.
At earlier epochs reductions of significantly over 50\% were seen. Note
that the
reduced SFRs continue to
offer similar peaks and troughs albeit at a lower overall normalization. 

\item The introduction of a self-gravity criterion to prevent catastrophic
star formation at high redshifts does not lead to significantly
different structure formation provided that the SFR
normalisation is set within reasonable bounds.

\item The treatment of the high density end of the SPH solver can produce
an enormous difference in the SFR. By reducing the neighbour search in the
high-density regions the SPH density becomes lower and
consequently so does the SFR. Also the self-gravity criterion is overcome
earlier since as the search radius is reduced the baryonic
cores have a higher weighting. 

\end{enumerate}

At
the resolution provided by the cosmological simulations 
hierarchical merging is not adequately modeled. Only a handful of
progenitor halos merge to form the major
disk system. That all of the simulations, including those with the
exceptional
feedback provided by temperature smoothing, produce an over-compact disk
is not cause for concern\-the radius of the disk is determined largely by
the
depth of the potential well. Due to the early stage at which the
simulations were stopped, it is difficult to comment on the
evolution of the majority of disk systems, which are expected to accrete a
large fraction of mass from $z\simeq1$ onwards. The one simulation that
was integrated to $z=0.5$ gave some surprising results. If feedback is
sufficiently strong to reduce the angular momentum problem for the
largest systems then there are insufficient halos at later times to
accrete on to the disk. Although it is a significant jump to go from this
result to the idea that feedback was stronger at higher redshifts it
certainly seems appealing. Silk (1998) discusses variations in
the
IMF over time.

One particularly important area that has not been examined is the effect
of
resolution. For star formation algorithms based upon density values, there
are many issues to be considered. In particular, in a hierarchical
cosmology, star formation will be higher and begin at earlier epochs with
increased resolution. The effect on feedback is less clear but by
increasing the mass resolution in simulations the escape velocity of the
first halos is reduced. Since the gas temperature following a feedback
event is not reduced (with increased resolution),  heated gas will tend to
orbit higher in the potential well and be more diffuse. It was noted that
in the simulations with energy smoothing (even those with a high SFR) the
formation of a dense central core was unavoidable. This might be partially
associated with the delay between the first star formation and the first
feedback. During this time it may be the case that gas can accumulate
above the mass threshold at which it can be blown out. A probablistic star
formation algorithm might improve this matter somewhat, although in the
limit $N\rightarrow\infty$ these algorithms should converge. 
However, it should be acknowledged that the dark matter halos in the sCDM
picture have a strong central concentration and that the core accumulation
may be a result of this.

As is widely known, core-halo angular momentum transport presents great
problems for the
sCDM picture. An important question is what is the extent of the
problem 
for merging halos of unequal mass? It is unclear whether the
limiting case of accreting a large number of low mass halos will lead to a
system that has not lost a significant proportion of its angular momentum.
Since the internal angular momentum of the final object is carried by the
orbital angular momentum of the progenitors, there is reason to believe
that it should be possible.  However, higher resolution SPH studies
(Thacker and Couchman, forthcoming)  indicate that it is almost impossible
to avoid the formation of large objects from mergers of smaller ones
without the angular momentum problem coming in to play -- there are too
many `medium sized' halos that collapse within the galaxy halos. It is
also possible to make insights without relying upon high resolution since
the power
spectrum relevant to galaxy formation is the approximately scale invariant
$P(k)\propto k^{-2}$ (cooling times are shorter for the smaller 
halos, and the power spectrum is tilted more toward $P(k)\propto k^{-3}$).
The simulations thus performed give a good
indication
of the properties of the low mass progenitors of galaxies -- they too
should suffer from an angular momentum deficit. However, since this is a
problem for the internal angular momentum and not the orbital, the effect
on the final object may not be a significant problem. In light of this
argument, it is seen that smooth infall in simulations occurs as a result
of a {\em lack of resolution}. Indeed the idea of smooth infall in sCDM is
largely a misnomer (at least without some kind of feedback mechanism).  In
simulations, the minimum mass scale effectively keeps the gas supported
against the collapse that would normally ensue given higher resolution.
Thus the disks that are formed in SPH simulations are a result of this
lack of numerical resolution, and ideally detailed convergence
studies
should be undertaken. A number of authors have already hinted at this
problem (Evrard \etal 1994; Weinberg, Hernquist \& Katz 1997, for
example), but as yet no
systematic attempt has been made to deal with it or
its effects. This problem is revisted in a paper in preparation.

The large variations in SFRs that may be
produced by small changes in the parameter space remain a significant
concern. The treatment of the
high density end of the SPH solver is of particular importance, since it
can strongly affect the SFR calculated, and further the treatment is not
performed in the same manner by all research groups. The development of a
standard test case for dynamical star formation algorithms is desireable, 
the
facilitating the comparison of different research results. Unfortunately,
it is far from clear what kind of a test case should be adopted. Simple
rotating cloud collapse models are not adequate since they provide no
representation of the hierarchical formation process or the effect of
tidal fields. Also, since some algorithms are grid-based and some are
particle-based, it may be necessary to adopt a suite of similar test
cases.

McGaugh (1998) presents a survey of a number of reasons why
sCDM is unable
to form low surface brightness (LSB) galaxies similar to those observed.
Nonetheless, it is still
instructive to compare the SFRs calculated with those from deep
observations of star forming galaxies. Studies of the global SFR, when
adjusted for dust extinction, suggest that there is no decline between
$z=1$ to 4.5. Consequently, the self-gravity criterion imposed,
causing an abrupt turn on of star formation at $z=4$, does not fit the
data, especially the observations of Chen \etal (1999) suggesting star
formation at $z\simeq7$.
This should not be over-interpreted since it is a result of a lack
of resolution, adding higher resolution would move the onset of star
formation to progressively earlier times. Turning off the criterion allows
star formation to proceed very early on, at $z=8.3$, but it also allows
star formation to occur in regions where the dark matter may still be
dynamically dominant, which is undesirable. Using the continuum UV flux,
the DEEP survey of the {\em Hubble Deep Field} derives SFRs from 0.14
\msolyr\ to 24.92 \msolyr\ for $q_0=0.05$ and $q_0=0.5$ values are 11
times lower (\cite{LOW97}).
Note
that these values are {\em not} corrected for dust extinction.  At $z=3$
the SFRs calculated (30-70 \msolyr) are higher than those found,
although the selected sample are categorized as ``large dwarf
spheroidals'' and, in contrast, at $z=3$ the simulated galaxy already has
a mass of
$7.2\times 10^{10}$ \msol. More recently estimates of SFRs derived from
H$\beta$ emission, for a sample of $z=3$ galaxies, have shown widely
diverging values compared to the estimate from the UV flux (with both
fluxes being uncorrected for dust extinction). The H$\beta$ values are
larger, by as much as a factor of 7, yielding SFRs in the range 20-270
\msolyr\ (\cite{PET98}). It is difficult to see how the models presented 
can be tuned
to produce SFRs in the region of 270 \msolyr\ since it would require an
SFR
normalization far beyond what we believe is the realistic parameter space.
A forthcoming paper examines the effect of higher spatial resolution
on the derived SFR.

\acknowledgements
The authors thank NATO for providing collaborative research
grant CRG 970081 which helped facilitate this research. Useful
discussions with
Drs Fabio Governato and Frazer Pearce are acknowledged. RJT was
supported by a
Dissertation Fellowship from the University of Alberta while this research
was conducted. HMPC ackowledges support from NSERC Canada.

\end{document}